\documentclass[journal,twoside,web]{ieeecolor}
\usepackage{generic}
\usepackage{cite}
\usepackage{amsmath,amssymb,amsfonts}
\usepackage{algorithmic}
\usepackage{graphicx}
\usepackage{textcomp}

\newtheorem{theorem}{Theorem}

\newtheorem{assumption}{Assumption}

\makeatletter
\let\NAT@parse\undefined
\makeatother
\usepackage{hyperref}
\usepackage{cleveref}
\hypersetup{colorlinks,linkcolor={black},citecolor={blue},urlcolor={blue}} 

\def\BibTeX{{\rm B\kern-.05em{\sc i\kern-.025em b}\kern-.08em
		T\kern-.1667em\lower.7ex\hbox{E}\kern-.125emX}}
\begin{document}
\title{Robust Sliding Mode Control of a Magnetic Levitation System: Continuous-Time and Discrete-Time Approaches}
\author{Pratik Vernekar and Vitthal Bandal
\thanks{P. Vernekar is with General Motors, Milford, MI 48380, USA (e-mail: pratik.vernekar@gmail.com). V. Bandal is with Government Polytechnic, Pune, Maharashtra 411005, India (e-mail: vitthalsbandal@gmail.com).}
}

\maketitle

\begin{abstract}
This paper presents three types of sliding mode controllers for a magnetic levitation system. First, a proportional-integral sliding mode controller (PI-SMC) is designed using a new switching surface and a proportional plus power rate reaching law. The PI-SMC is more robust than a feedback linearization controller in the presence of mismatched uncertainties and outperforms the SMC schemes reported recently in the literature in terms of the convergence rate and settling time. Next, to reduce the chattering phenomenon in the PI-SMC, a state feedback-based discrete-time SMC algorithm is developed. However, the disturbance rejection ability is compromised to some extent. Furthermore, to improve the robustness without compromising the chattering reduction benefits of the discrete-time SMC, mismatched uncertainties like sensor noise and track input disturbance are incorporated in a robust discrete-time SMC design using multirate output feedback (MROF). With this technique, it is possible to realize the effect of a full-state feedback controller without incurring the complexity of a dynamic controller or an additional discrete-time observer. Also, the MROF-based discrete-time SMC strategy can stabilize the magnetic levitation system with excellent dynamic and steady-state performance with superior robustness in the presence of mismatched uncertainties. The stability of the closed-loop system under the proposed controllers is proved by using the Lyapunov stability theory. The simulation results and analytical comparisons demonstrate the effectiveness and robustness of the proposed control schemes.
\end{abstract}


\section{Introduction}\label{sec1}

\IEEEPARstart{M}{agnetic} levitation systems are widely used in many engineering applications due to the significant advantages of contact-less motion, no lubrication, no pollution, and long service life. Some of the typical applications are high-speed maglev trains, frictionless magnetic bearings, and future transportation systems like the Hyperloop \cite{b1}. The Hyperloop, which uses magnetic levitation, would be a big revolution in transportation technology that is faster, safer, cheaper, and greener, which would help in solving the challenges of climate change. Magnetic levitation systems are usually open-loop unstable and are described by highly nonlinear differential equations. Also, the uncertainties in these systems may not always be on the same channel as the control input or satisfy the matching conditions. Therefore, it is necessary to design a high-performance feedback controller that not only stabilizes the system but is also robust to mismatched uncertainties.

Many advanced control methods have been reported in the literature so far for effectively controlling magnetic levitation systems. First of all, several nonlinear control techniques have been developed \cite{b2,b3,b4,b5}. Hajjaji \emph{et al.} developed and validated a feedback linearization controller on an experimental prototype of the magnetic levitation system \cite{b2}. An adaptive robust nonlinear controller via the backstepping design approach is proposed in \cite{b3} to solve the position tracking problem of a magnetic levitation system in the presence of parametric uncertainties. Yang \emph{et al.} presented a robust nonlinear controller using the dynamic surface control technique that is superior to the conventional backstepping-technique-based nonlinear controller \cite{b4}. This controller showed excellent position-tracking performance in the presence of modeling errors. In \cite{b5}, a nonlinear model predictive control (MPC) scheme based on a real-time gradient algorithm is presented. This controller showed better performance than a linear MPC in the presence of measurement noise and model uncertainties. Several sliding mode control (SMC) strategies have also been applied to magnetic levitation systems \cite{b6,b7,b8,b9,b10,b11,b12}.

Sliding mode control (SMC) is a popular method because the discontinuous nature of the control action results in outstanding robustness features such as insensitivity to parametric variations and rejection of external disturbances \cite{b13,b14}. SMC schemes of the static and dynamic types are presented in \cite{b6} for the control of a magnetic levitation system. The adaptive fast terminal SMC (AFTSMC) method presented in \cite{b7} shows good tracking results and finite-time convergence to the sliding surface. Pan \emph{et al.} proposed an improved double exponential reaching law-based integral SMC (DPRL-I-SMC) algorithm for a magnetic suspension system \cite{b8}. This controller performed well under a sinusoidal disturbance and significantly reduced the chattering of the system. In \cite{b9}, a fractional-order SMC (FOSMC) is proposed. This controller performed better than a traditional SMC in tracking accuracy, speed of response, and control energy. An adaptive terminal SMC technique based on disturbance compensation has been developed by Wang \emph{et al.} \cite{b10}. By adaptively tuning the switching gain value, the controller showed excellent dynamic and steady-state performance. Also, Wang \emph{et al.} incorporated a reduced-order generalized proportional-integral observer in the SMC design to reduce the chattering phenomenon without sacrificing the time-varying disturbance rejection ability \cite{b11}. However, the above algorithms do not completely address the robustness of the controllers in the presence of matched and mismatched uncertainties in the system. The overall robustness of the system could be significantly improved in the proposed methods \cite{b2,b3,b4,b5,b6,b7,b8,b9,b10,b11,b12}.

The main emphasis of this paper is to develop robust SMC algorithms in the continuous- and discrete-time domain for a magnetic levitation system, in which matched and mismatched uncertainties of the system are taken into account in the controller design. In addition to the robustness of the controllers, this paper also addresses the phenomenon of chattering present in the continuous-time SMC methods. Building upon the preliminary work reported in \cite{b15, b16}, this study investigates the performance of three sliding mode controllers.

We first propose a continuous-time proportional-integral sliding mode control (PI-SMC) algorithm using a new sliding surface and a proportional plus power rate reaching law. The proposed PI-SMC has the advantages of a feedback linearization controller of being insensitive to deviations from the nominal operating point and can be directly applied to nonlinear dynamics without linear approximations. Also, the proportional-integral sliding surface drives the system states to the desired equilibrium with zero steady-state error. With the combined advantages of a feedback linearization controller and a proportional-integral sliding surface, the proposed PI-SMC performs better than the SMC algorithms \cite{b6,b7,b8,b9,b10,b11,b12} designed so far for a magnetic levitation system in the presence of mismatched uncertainties. The robustness and stability of the algorithm are analytically proved using the Lyapunov stability theory and verified in simulation through disturbance analysis. Also, the proportional and power rate portions of the reaching law ensure that the approach speed is fast when the system states are far away or close to the sliding surface. In addition to the superior robustness features, the PI-SMC shows a better convergence rate than the AFTSMC \cite{b7}, improved DPRL-I-SMC \cite{b8}, and FOSMC \cite{b9} control schemes reported recently in the literature. Thus, the PI-SMC shows good dynamic performance with fast convergence to the sliding surface and finally to the equilibrium point. However, the above algorithm has the drawbacks of control input chattering. 

To address the chattering issue in the PI-SMC, we propose a state feedback-based discrete-time SMC method using the improved discrete-time reaching law \cite{b17}. Using the boundary layer concept by appropriately designing the width of the quasi-sliding mode (QSM) band, the chattering present in the PI-SMC is significantly reduced in the discrete-time SMC. For the design of the discrete-time SMC, we first derive the nonlinear model of the magnetic levitation system in the discrete-time domain. To the best knowledge of the authors, robust discrete-time SMC algorithms, which consider matched and mismatched uncertainties of the system, have not been designed so far for a magnetic levitation system. We propose two unique methods for a robust state feedback-based discrete-time SMC algorithm design, such that the closed-loop system is stable in the presence of matched uncertainties. However, due to the QSM motion of the state trajectory, the robustness property present in the PI-SMC is lost to some extent in the state feedback-based discrete-time SMC.

To improve the robustness of the state feedback-based discrete-time SMC, we propose a robust discrete-time SMC strategy using multirate output feedback (MROF) \cite{b18,b19,b20} for a magnetic levitation system. In MROF, the system output is sampled at a faster rate than the control input, and only the system outputs and the past control inputs are used to compute the control input making this technique more practical for a magnetic levitation system compared to state feedback-based methods, in which separate observers are required to estimate the velocity of the ferromagnetic ball and other state variables.

The SMC strategies reported in the literature so far for a magnetic levitation system are either based on state feedback \cite{b6,b7,b8,b9} or employ complex disturbance observers \cite{b10,b11,b12} to estimate the unmeasurable states. Also, the system parameters of practical systems are not always accurately known. In this case, the system states and parameters are estimated using sophisticated gradient-based iterative algorithms, data filtering techniques, and system identification methods \cite{b21,b22,b23,b24,b25,b26,b27}. The proposed MROF-based controller consists of a simple and easy-to-implement MROF-based state estimator and doesn't require a complex disturbance observer to reject the external disturbances or an intricate system identification method to estimate the system parameters. The disturbance compensation components, the state estimator, and the MROF-based controller gains can be easily designed and tuned to improve the overall robustness of the system. Thus, using the MROF-based technique, it is possible to realize the effect of full-state feedback without incurring the complexity of a dynamic controller. Furthermore, the controller is designed considering mismatched uncertainties like sensor noise and track input disturbance in the system dynamics. The proposed MROF-based strategy can achieve QSM motion and stabilize the system in the presence of mismatched uncertainties with reduced chattering.

The main contributions of this paper are as follows:
\newline 1) The proposed PI-SMC algorithm has the combined advantages of a feedback linearization controller, a proportional-integral sliding surface, and a proportional plus power rate reaching law. As a result, the PI-SMC performs better than the SMC techniques reported in the literature in the presence of mismatched uncertainties. A comparison is made between the proposed PI-SMC and the AFTSMC \cite{b7}, DPRL-I-SMC \cite{b8}, and FOSMC \cite{b9} control schemes for the convergence rate and settling time to demonstrate the effectiveness of the algorithm.
\newline 2) We propose two unique methods for a robust state feedback-based discrete-time SMC design to eliminate the chattering phenomenon and stabilize the closed-loop system in the presence of matched uncertainties.
\newline 3) The proposed MROF-based discrete-time SMC strategy has the following unique characteristics which distinguish it from existing SMC methods: a) Is robust to both matched as well as mismatched uncertainties in the system, b) Uses only the system outputs and the past control inputs to compute the control input, c) Doesn't require complex disturbance observers or state-of-the-art system identification methods to estimate the unmeasurable states and parameters, d) Mismatched uncertainties like measurement sensor noise and track input disturbance are considered in the controller design, e) The design parameters in the control law, sliding surface, disturbance compensation components, and the state estimator can be appropriately tuned to achieve the desired level of robustness, improve the convergence rate, reduce chattering, and decrease the control effort and energy, f) Finally, using this strategy it is possible to realize the effect of full-state feedback with superior robustness features.
\newline 4) A rigorous finite-time convergence and stability analysis of the proposed controllers is done using the Lyapunov stability theory. We have proved the attractiveness and invariance of the switching surface during reaching and sliding modes and the overall closed-loop stability of the system for the proposed control schemes. Comprehensive stability analysis of the proposed controllers is one of the main highlights of this paper.
\newline 5) A comparative analysis of the proposed control schemes is done using four well-known performance criteria. Also, the proposed control schemes are qualitatively compared with other SMC methods reported in the literature.

The rest of the paper is organized as follows. In Section \ref{sec2}, the model of the magnetic levitation system is presented. Section \ref{sec3} presents the design of the continuous-time PI-SMC technique. In Section \ref{sec4}, the state feedback-based discrete-time SMC method is described. The multirate output feedback-based discrete-time SMC strategy is proposed in Section \ref{sec5}. In Section \ref{sec6}, the simulation results and analytical comparisons of the three controllers are reported. Finally, Section \ref{sec7} concludes the research findings.

\section{Model of the Magnetic Levitation System}\label{sec2}

The magnetic levitation system shown in Fig. \ref{fig1} consists of a controller, a power amplifier, a position sensor, an electromagnetic coil, and a ferromagnetic ball suspended in a magnetic field generated by an electromagnet. The electromagnet exerts an attractive force, $F(i, p)$, on the ferromagnetic ball to counteract the gravitational force, $mg_c$, to keep the ball at the desired height. The dynamic model of the magnetic levitation system can be expressed as 
\begin{gather}
	\begin{aligned}	
		&\dfrac{dp}{dt} = v, \ \ Ri + \dfrac{d(L(p)i)}{dt} = e, \ \ m\dfrac{dv}{dt} = mg_c - F(i, p),\\
		&F(i, p) = Q\bigg{(}\dfrac{i}{p}\bigg{)}^2, \ \ Q = \dfrac{\mu _{0}AN^2}{4}, \ \ L(p) = L_1 + \dfrac{2Q}{p},
	\end{aligned}
	\label{eq1}
\end{gather}
in which $p$ is the ball position, $v$ is the ball velocity, $i$ is the coil current, $m$ is the ball mass, $e$ is the input voltage, $R$ is the coil resistance, $L$ is the electromagnetic coil inductance, which is a nonlinear function of the ball position $p$, and is dependent on system parameter $L_1$, $g_c$ is the gravitational constant, $Q$ is the magnetic force constant, $\mu _{0}$ is the vacuum permeability, $A$ is the magnetic permeability area, and $N$ is the coil turn. Let us define the state vector as $x = (x_1 \ \ x_2 \ \ x_3)^T = (p \ \ v \ \ i)^T$, the control input as $u = e$, and the output as $y = x_1 = p$. Thus, the dynamic model \eqref{eq1} of the magnetic levitation system can be written in the nonlinear state-space form as follows:
\begin{figure}[!t]
	\centerline{\includegraphics[width=\columnwidth]{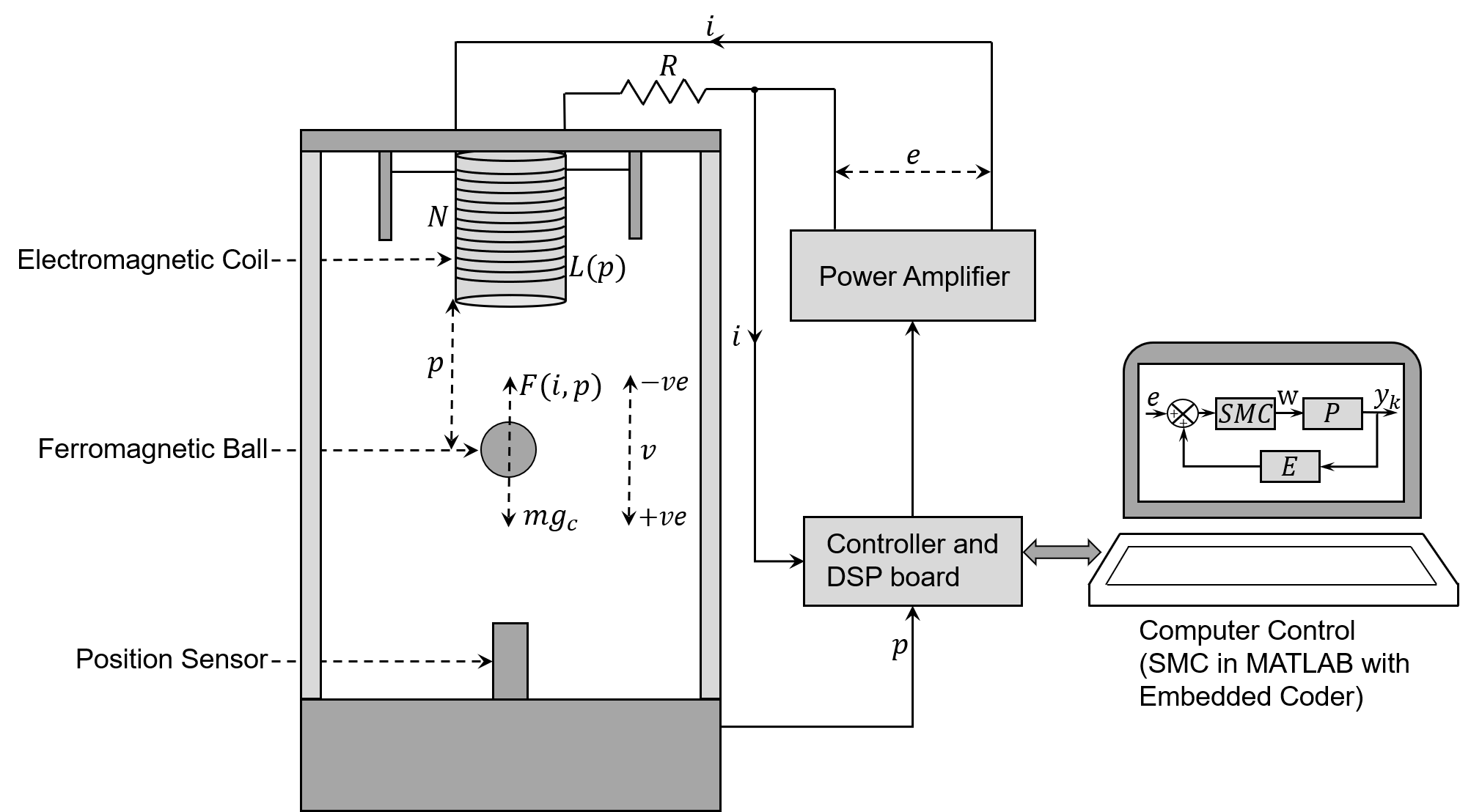}}
	\caption{Schematic of the magnetic levitation system.}
	\label{fig1}
\end{figure}
\begin{table}[!h]
	\setlength{\tabcolsep}{4pt}
	\small
	\centering
	\caption{Parameters of the magnetic levitation system}
	\begin{tabular}{|c|c|c|}
		\hline
		Symbol& 
		Parameter& 
		Value\\
		\hline
		$R$&
		Coil resistance&
		28.7 $\Omega$ \\ \hline
		$L_1$& 
		Coil inductance& 
		0.65 H \\ \hline
		$g_c$& 
		Gravitational constant& 
		9.81 m/s$^2$ \\ \hline
		$m$& 
		Ball mass& 
		11.87 $\times$ 10$^{-3}$ kg \\ \hline
		$\mu _{0}$& 
		Vacuum permeability& 
		2.125 $\times$ 10$^{-7}$ H/m \\ \hline
		$A$& 
		Magnetic permeability area& 
		8$\pi$ $\times$ 10$^{-4}$ m$^2$ \\ \hline
		$N$& 
		Coil turn& 
		1024  \\ \hline
		$Q$&         
		Magnetic force constant& 
		1.4 $\times 10^{-4}$ \\ \hline
	\end{tabular}
	\label{tab1}
\end{table}
\begin{equation}
	\begin{aligned}	
		&\dot{x} = f(x) + g(x)u + d(x),\ \ \ y = h(x),\\
	\end{aligned}
	\label{eq2}
\end{equation}
in which $x\in \mathbb{R}^3$, $u\in\mathbb{R}$, $y\in\mathbb{R}$, and 
\begin{equation}
	\begin{aligned}	
		&f(x) = \begin{pmatrix} f_1(x)\\ f_2(x)\\ f_3(x) \end{pmatrix} = \begin{pmatrix} x_2\\ g_c - \dfrac{Q}{m}\bigg{(}\dfrac{x_3}{x_1}\bigg{)}^2\\ 
			-\dfrac{R}{L(x_1)}x_3 + \dfrac{2Q}{L(x_1)}\bigg{(}\dfrac{x_2x_3}{x_1^2}\bigg{)}\end{pmatrix},\\  
		&g(x) = \begin{pmatrix} 0\\ 0\\ \dfrac{1}{L(x_1)} \end{pmatrix}, \ \ \ d(x) = \begin{pmatrix} d_1(x)\\ d_2(x)\\ d_3(x) \end{pmatrix}, \ \ \ h(x) = x_1.\\
	\end{aligned}
	\label{eq3}
\end{equation}

In \eqref{eq3}, $f(x)$ and $g(x)$ are smooth continuous functions, and $d(x)$ is the total disturbance which is comprised of parametric and model uncertainties, unknown internal dynamics, and unmeasurable external disturbances and nonlinearities. The disturbances $d_1(x)$ and $d_2(x)$ are mismatched, while the disturbance $d_3(x)$ is matched with the control input $u$. The objective of the control schemes is to drive the states $x_1$, $x_2$, and $x_3$, to their desired steady-state values $x_{1d}$, $x_{2d}$, and $x_{3d}$, respectively. From \eqref{eq3}, the equilibrium point for the system is $x_d = (x_{1d} \ \ 0 \ \ x_{3d})^T$, in which $x_{3d} = x_{1d}\sqrt{\frac{g_c}{Q}m}$. Some of the experimental prototypes of a magnetic levitation system are given in \cite{b5,b9,b10,b11}. In this study, the system parameters given in Table \ref{tab1} are derived from \cite{b6}.

\section{Proportional-Integral Sliding Mode Control}\label{sec3}
In this section, we propose a continuous-time PI-SMC algorithm using a new sliding surface and a proportional plus power rate reaching law. Using PI-SMC is beneficial because it is robust to systems with mismatched uncertainties. The sliding surface equation consists of a linear term plus an integral term and allows direct use of the pole placement technique. The controller design consists of the following steps: a) Nonlinear feedback linearization of the magnetic levitation system model, b) Proportional-integral sliding surface design, and c) Control law formulation based on a proportional plus power rate reaching law.

\subsection{Controller Design}\label{sec3subsec1}
The first step in the design of the PI-SMC is feedback linearization of the magnetic levitation system model. The nonlinear model of the magnetic levitation system given in \eqref{eq2} and \eqref{eq3} is linearized using the feedback linearization technique \cite{b28}. For the nonlinear model of the magnetic levitation system, we assume that the vector fields $f: D\rightarrow\mathbb{R}^3$, $g: D\rightarrow\mathbb{R}^3$, and the readout map $h: D\rightarrow\mathbb{R}$ are smooth in the domain $D\subset\mathbb{R}^3$, i.e., their partial derivatives with respect to $x$ of any order exist and are continuous in $D$. Let us define the nonlinear change of variables $z = (z_1 \ \ z_2 \ \ z_3)^T$ as
\begin{equation}
	\begin{aligned}	
		&z = T(x) = \begin{pmatrix} h(x) - x_{1d}\\ L_fh(x)\\ L^2_fh(x) \end{pmatrix} = \begin{pmatrix} x_1 - x_{1d}\\ x_2\\ g_c - \dfrac{Q}{m}\bigg{(}\dfrac{x_3}{x_1}\bigg{)}^2\end{pmatrix},\\  
	\end{aligned}
	\label{eq4}
\end{equation}
in which $L_fh(x)$ is the Lie derivative of the function $h(x)$ with respect to the vector field $f(x)$ and $L^2_fh(x) = L_f(L_fh(x))$. The state-space equations in the new coordinates are as follows:
\begin{equation}
	\begin{aligned}	
		&\dot{z} =\\ &\begin{pmatrix} \dot{x}_1 \\ \dot{x}_2\\ L^3_fh(x) + L_g(L^2_fh(x))u\end{pmatrix} = \begin{pmatrix} z_2 + d_1(z)\\ z_3 + d_2(z)\\ \alpha(z) + \beta(z)u + d_3(z)\end{pmatrix},\\  
	\end{aligned}
	\label{eq5}
\end{equation}
in which $d(z) = (d_1(z) \ d_2(z) \ d_3(z))^T$ is the disturbance and 
\begin{equation}
	\begin{aligned}	
		&\alpha(z) = 2(g_c-z_3)\bigg{(}\bigg{(}1 - \dfrac{2Q}{L(z_1 + x_{1d})}\bigg{)}\dfrac{z_2}{(z_1 + x_{1d})} + \dfrac{R}{L}\bigg{)},\\
		&\beta(z) = -\dfrac{2}{L(z_1 + x_{1d})}\sqrt{\dfrac{Q}{m}(g_c - z_3)}.\\		  
	\end{aligned}
	\label{eq6}
\end{equation} 
To cancel the nonlinearities in \eqref{eq5}, the smooth outer-loop feedback controller $u$ is designed as follows: $u = \beta^{-1}(z)(-\alpha(z) + w)$, in which $w$ is the inner-loop feedback controller. The time subscript $t$ has been omitted in the equations so far for brevity. Thus, the nonlinear model of the magnetic levitation system \eqref{eq2}, can be written in a linear Brunovsky form as
\begin{equation}
	\begin{aligned}	
		&\dot{z}(t) = Az(t) + Bw(t) + d(z, t),\ \ \
		y(t) = Cz(t),\\
	\end{aligned}
	\label{eq8}
\end{equation}
in which
\begin{equation}
	\begin{aligned}	
		&A = \begin{pmatrix} 0 & 1 & 0\\ 0 & 0 & 1\\ 0 & 0 & 0\end{pmatrix}, \ \ \ B = \begin{pmatrix} 0\\ 0\\ 1\end{pmatrix}, \ \ \ C = (1 \ \ 0 \ \ 0).\\  
	\end{aligned}
	\label{eq9}
\end{equation}

The second step in the design of the PI-SMC is design of the switching surface $s(t)$, which is as follows: 
\begin{equation}
	\begin{aligned}	
		&s(t) = M^Tz(t)-\int_0^tM^T(A+BK)z(\tau)d\tau,\\  
	\end{aligned}
	\label{eq10}
\end{equation} 
in which $M^T  = (m_1 \ \ m_2 \ \ m_3) \in\mathbb{R}^{1\times3}$ and $K\in\mathbb{R}^{1\times3}$ are constant matrices. If the matrices $K$ and $M$ are designed such that $H = A + BK$ is Hurwitz, and $M^TB = m_3 \neq 0$ is invertible, the states of the closed-loop system will enter the switching surface in a finite time and will be zero asymptotically. Using \eqref{eq9}, simple calculations show that
\begin{equation}
	\begin{aligned}	
		M^T(A+BK) = & \ (m_1 \ \ m_2 \ \ m_3) \begin{pmatrix} 0 & 1 & 0\\ 0 & 0 & 1\\ k_1 & k_2 & k_3\end{pmatrix},\\
		= & \  (m_3k_1 \ \ m_1 + m_3k_2 \ \ m_2 + m_3k_3).\\
	\end{aligned}
	\label{eq10a}
\end{equation} 

The third step in the design of the PI-SMC is formulation of the inner-loop feedback controller $w(t)$ using the following proportional plus power rate reaching law: 
\begin{equation}
	\begin{aligned}	
		&\dot{s}(t) = -k_4s(t) - k_0\lvert s(t)\rvert^{\alpha}\mathrm{sgn}(s(t)) - k_5\mathrm{sgn}(s(t)),\\
	\end{aligned}
	\label{eq11}
\end{equation}   
in which $k_4 > 0$, $k_0 > 0$, and $k_5 > 0$ are the rates at which the system state trajectory approaches the switching surface when it is far away or close to the switching surface, and $0 < \alpha < 1$. By differentiating \eqref{eq10}, we get $\dot{s}(t) = M^T\dot{z}(t) - M^T(A+BK)z(t)$. Substituting \eqref{eq8} and \eqref{eq11} in the above equation, and solving for the control input $w(t)$, yields
\begin{equation}
	\begin{aligned}	
		w(t) = & \ Kz(t) - k_0(M^TB)^{-1}\lvert s(t)\rvert^{\alpha}\mathrm{sgn}(s(t)) \\
		& - k_4(M^TB)^{-1}s(t) - k_5(M^TB)^{-1}\mathrm{sgn}(s(t)) \\
		& - (M^TB)^{-1}M^Td(z, t). 
	\end{aligned}
	\label{eq13}
\end{equation}
However, the disturbance $d(z, t)$ cannot be measured, so we ignore the $d(z, t)$ term in the inner-loop feedback control law \eqref{eq13}. Using the relation $u = \beta^{-1}(z)(-\alpha(z) + w)$, the outer-loop PI-SMC input $u(t)$ is computed as follows:
\begin{equation}
	\begin{aligned}	
		u(t) = & \ \beta^{-1}(z(t))[-\alpha(z(t)) + Kz(t)\\
		& - k_4(M^TB)^{-1}s(t) - k_5(M^TB)^{-1}\mathrm{sgn}(s(t))\\ 
		& - k_0(M^TB)^{-1}\lvert s(t)\rvert^{\alpha}\mathrm{sgn}(s(t))].  
	\end{aligned}
	\label{eq15}
\end{equation}

\subsection{Stability Analysis}\label{sec3subsec2}
In this section, we prove that the outer-loop PI-SMC feedback controller $u(t)$ asymptotically stabilizes the nonlinear model of the magnetic levitation system \eqref{eq2}. Instead of using system \eqref{eq2} in the original coordinates, we use system \eqref{eq5} in the new coordinates for the stability analysis.

\begin{assumption}\label{assumption1}
	The disturbance $d(x, t)$ in system \eqref{eq2} and thus the disturbance $d(z, t)$ in system \eqref{eq5} is continuous and bounded, and is defined by $D_i = \sup_{t > 0} \lvert d_i(z, t)\rvert$ for $i$ = 1, 2, 3, i.e., there exist constants $D_1 > 0$, $D_2 > 0$, and $D_3 > 0$, such that $\lvert d_1(z, t)\rvert \leq D_1$, $\lvert d_2(z, t)\rvert \leq D_2$, and $\lvert d_3(z, t)\rvert \leq D_3$, $\forall t > 0$. Also, the derivative of the disturbance in system \eqref{eq5} is bounded and satisfies $\lim\limits_{t \to \infty} \dot{d}_i(z, t) = 0$ for $i$ = 1, 2, 3. 
\end{assumption}

\begin{theorem}\label{thm1}
	Assume that the nonlinear model of the magnetic levitation system \eqref{eq5} satisfies Assumption \ref{assumption1}. Then the closed-loop system under the outer-loop PI-SMC feedback law \eqref{eq15}, in which the sliding surface $s(t)$ is as given in \eqref{eq10}, is asymptotically stable if the control gains satisfy the following conditions: $k_5 \geq m_1D_1 + m_2D_2 + m_3D_3 + \eta$, $\eta > 0$, $k_4 > 0$, $k_0 > 0$, $0 < \alpha < 1$, the gain $K = (k_1 \ \ k_2 \ \ k_3)$ is designed such that the matrix $A + BK$ is Hurwitz i.e., $\lambda_{\mathrm{max}} (A + BK) < 0$, and $M^T = (m_1 \ \ m_2 \ \ m_3)$ is chosen such that the polynomial
	$p(s) = m_1s^2 + m_2s + m_3$ is a Hurwitz polynomial.  
\end{theorem}

\textit{Proof:} The analysis is carried out by using $V(s(t)) = \frac{1}{2}s^T(t)s(t)$ as a Lyapunov function candidate. By differentiating $V(s(t))$ along the system state trajectory, we get $\dot{V}(s(t)) = s(t)\dot{s}(t) = s(t)[M^T\dot{z}(t) - M^T(A+BK)z(t)]$. Substituting the nonlinear state equation \eqref{eq5} into the above equation, yields
	\begin{equation}
		\begin{aligned}	
			\dot{V}(s(t)) = s(t)[&m_1z_2(t) + m_1d_1(z, t) + m_3\beta(z(t))u(t)\\
			&+ m_2d_2(z, t) + m_3\alpha(z(t)) + m_2z_3(t)\\
			&+ m_3d_3(z, t) - M^T(A + BK)z(t)].\\  
		\end{aligned}
		\label{eq21}
	\end{equation}
	By substituting \eqref{eq10a} and the outer-loop PI-SMC control law \eqref{eq15} into \eqref{eq21}, we get
	\begin{equation}
		\begin{aligned}	
			\dot{V}(s(t)) = & \ s(t)[m_3\alpha(z(t)) + m_1d_1(z, t) + m_2d_2(z, t)\\
			& + m_1z_2(t) - m_3k_1z_1(t)  + m_2z_3(t) - m_3\alpha(z(t))\\ 
			& + m_3k_1z_1(t) + m_3k_2z_2(t) - k_4m_3(M^TB)^{-1}s(t)\\
			& + m_3k_3z_3(t)  - k_5m_3(M^TB)^{-1}\mathrm{sgn}(s(t))\\
			& - k_0m_3(M^TB)^{-1}\lvert s(t)\rvert^{\alpha}\mathrm{sgn}(s(t)) + m_3d_3(z, t)\\ 
			&  - (m_1 + m_3k_2)z_2(t) - (m_2 + m_3k_3)z_3(t)].\\ 		                      
		\end{aligned}
		\label{eq22}
	\end{equation}
	Using the relation $M^TB = m_3$ and simplifying \eqref{eq22}, yields $\dot{V}(s(t)) = s(t)[m_1d_1(z, t) + m_2d_2(z, t) + m_3d_3(z, t) - k_4s(t) - k_5\mathrm{sgn}(s(t)) - k_0\lvert s(t)\rvert^{\alpha}\mathrm{sgn}(s(t))]$. By substituting $\lvert s(t)\rvert = s(t)\mathrm{sgn}(s(t))$ into the above equation, we get
	\begin{equation}
		\begin{aligned}	
			\dot{V}(s(t)) = &-k_4[s(t)]^2 - k_0\lvert s(t)\rvert^{\alpha}\lvert s(t)\rvert + m_1d_1(z, t)s(t)\\
			& + m_2d_2(z, t)s(t) + m_3d_3(z, t)s(t) - k_5\lvert s(t)\rvert\\
			\leq & \ [-k_4\lvert s(t)\rvert - k_0\lvert s(t)\rvert^{\alpha} - k_5 + m_1\lvert d_1(z, t)\rvert\\ 
			&+ m_2\lvert d_2(z, t)\rvert + m_3\lvert d_3(z, t)\rvert]\lvert s(t)\rvert\\
			\leq & \ [-k_4\lvert s(t)\rvert - k_0\lvert s(t)\rvert^{\alpha} - k_5\\ 
			&+ m_1D_1 + m_2D_2 + m_3D_3]\lvert s(t)\rvert.		                      
		\end{aligned}
		\label{eq24}
	\end{equation}
	Substituting the relationship $k_5 \geq m_1D_1 + m_2D_2 + m_3D_3 + \eta$ as given in Theorem \ref{thm1} into \eqref{eq24}, yields
	\begin{equation} 
		\begin{aligned}	
			&\dot{V}(s(t)) \leq  \ [-k_4\lvert s(t)\rvert - k_0\lvert s(t)\rvert^{\alpha} - \eta]\lvert s(t)\rvert \leq 0.\\		                      
		\end{aligned}
		\label{eq25}
	\end{equation}
	From \eqref{eq25}, we have proved that $\dot{V}(s(t)) < 0$ for $s(t) \neq 0$. This guarantees that the state trajectories for the nonlinear system \eqref{eq5} reach the sliding surface, $s(t) = 0$, in a finite time under the proposed outer-loop PI-SMC feedback law provided the controller gains satisfy the conditions in Theorem \ref{thm1}. Thus, the closed-loop system \eqref{eq5}, under the application of the outer-loop feedback controller $u(t)$, is asymptotically stable. This implies that the overall nonlinear system \eqref{eq2} in the original coordinates is also asymptotically stable. Therefore, we can conclude that the PI-SMC algorithm given by \eqref{eq15} guarantees the asymptotic convergence of the system states $x_1$, $x_2$, and $x_3$ to their desired equilibrium values.

\section{State Feedback-Based Discrete-Time Sliding Mode Control}\label{sec4}
In this section, we present a state feedback-based discrete-time SMC algorithm for a magnetic levitation system. We use the boundary layer concept by appropriately designing the width of the QSM band in the discrete-time SMC method to reduce chattering present in the continuous-time PI-SMC. In the state feedback-based discrete-time SMC technique, the control input is calculated once in every sampling interval and is held constant during this period. In this method, the system state trajectory will move monotonically towards the switching plane and cross it in a finite time. The controller design consists of the following steps: a) Discretization of the magnetic levitation system model, b) Switching surface design, and c) Control law formulation based on the discrete-time reaching law.

\subsection{Controller Design}\label{sec4subsec1}
The nonlinear model of the magnetic levitation system \eqref{eq5} is finitely discretizable at the order $N = 3$ under the coordinate transformation given in \eqref{eq4}. This implies that any trajectory of the nonlinear system \eqref{eq5} can be computed explicitly by a finite number ($N = 3$) of state derivatives. For a sampling time of $\tau$ seconds, the discrete-time representation of the system \eqref{eq5}, using the Taylor series approximation, can be written as
\begin{equation}
	\begin{aligned}	
		z_1(k+1) =& \ z_1(k) + \tau z_2(k) + \dfrac{\tau^2}{2}z_3(k) + d_1(z, u),\\
		z_2(k+1) =& \ z_2(k) + \tau z_3(k) + \dfrac{\tau^2}{2}\vartheta_\tau(z, u) + d_2(z, u),\\
		z_3(k+1) =& \ z_3(k) + \tau \vartheta_\tau(z, u) + \dfrac{\tau^2}{2}\kappa_\tau(z, u)\vartheta_\tau(z, u)\\ & + d_3(z, u),\\
	\end{aligned}
	\label{eq26a}
\end{equation}
in which $\vartheta_\tau(z, u) = \alpha_\tau(z) + \beta_\tau(z)u$, $\kappa_\tau(z, u) = \frac{\partial \vartheta_\tau(z, u)}{\partial z}$, $\alpha_\tau(z)$ and $\beta_\tau(z)$ are the discrete-time representations of $\alpha(z)$ and $\beta(z)$, respectively. The subscript $k\tau = k$ has been omitted in $\vartheta_\tau(z, u)$, $\kappa_\tau(z, u)$, $d_1(z, u)$, $d_2(z, u)$, and $d_3(z, u)$ for brevity. Also, the control input $u$ is piecewise constant over the sampling interval $\tau$, i.e., the following zero-order hold (ZOH) assumption holds: $u(t) = u(k\tau) = u(k)$ = constant for $k\tau \leq t < (k+1)\tau$. The equivalent linear discrete-time representation of the nonlinear model \eqref{eq26a} is as follows:
\begin{equation}
	\begin{aligned}	
		z(k+1) &= \Phi_\tau z(k) + \Gamma_\tau w(k) + D_\tau d(k),\\
		y(k) &= C_\tau z(k),\\
	\end{aligned}
	\label{eq26}
\end{equation}
in which the state $z(k)\in\mathbb{R}^3$, the input $w(k)\in \mathbb{R}$, and the output $y(k)\in \mathbb{R}$. Also, $\Phi_\tau = e^{A\tau} \in \mathbb{R}^{3\times3}$, $\Gamma_\tau = A^{-1}(e^{A\tau} - I) \in \mathbb{R}^{3\times1}$, $D_\tau \in \mathbb{R}^{3\times1}$, and $C_\tau \in \mathbb{R}^{1\times3}$, and the disturbance $d(k) \in \mathbb{R}$ is bounded with a known constant bound. When the disturbance $ d(k)$ is matched with the control input $w(k)$, the disturbance matrix $D_\tau = \Gamma_\tau$. For a sampling time $\tau = 0.1$ s, the discrete-time system matrices are as follows:
\begin{gather}
	\begin{aligned}	
		&\Phi_\tau = \begin{pmatrix} 1 & 0.1 & 0.005\\ 0 & 1 & 0.1\\ 0 & 0 & 1\end{pmatrix}, \ \ \ \Gamma_\tau = \begin{pmatrix} 0.0002\\ 0.005\\ 0.1\end{pmatrix}.\\  
	\end{aligned}
	\label{eq27}
\end{gather}

After discretization of the system model, the switching surface, $s(k) = M^Tz(k)$, in which $s(k)\in \mathbb{R}$ and $M^T\in\mathbb{R}^{1\times3}$, is designed next. We design the switching surface $s(k)$ such that the system dynamics are stable when confined to the QSM band. The vector $M$ satisfies the following relation: $M^T\Gamma_\tau = M^T A^{-1}(e^{A\tau} - I) \neq 0$. The state trajectory for the discrete-time model of the magnetic levitation system \eqref{eq26} stays within the QSM band if the switching surface $s(k)$, for any $k \geq 0$, satisfies the following conditions \cite{b17}:
\begin{align}	
	&s(k) > \xi \Rightarrow 0 < s(k+1) < s(k), \label{eq28a}\\ 
	&s(k) < -\xi \Rightarrow 0 > s(k+1) > s(k),\label{eq28b}\\ 
	&\lvert s(k)\rvert \leq \xi \Rightarrow \lvert s(k+1)\rvert \leq \xi, \label{eq28c}			
\end{align}
in which $\Theta = \{z(k) \lvert -\xi \leq s(k) \leq \xi\}$ is the QSM band with a width of $2\xi$. If $\xi = 0$, the QSM is called an ideal QSM. Equations \eqref{eq28a}--\eqref{eq28c} guarantee the attractiveness of the QSM $2\xi$-band and ensure that the discrete-time system \eqref{eq26} satisfies the reaching condition of QSM motion.

Next step in the design of the discrete-time SMC is formulation of the controller $w(k)$ using the discrete-time reaching law as given below:
\begin{equation}
	\begin{aligned}	
		s(k+1) - s(k) = & -q\tau s(k) - \varepsilon \tau \mathrm{sgn}(s(k))\\ 
		& + \tilde{d}(k) - d_m - d_s\mathrm{sgn}(s(k)),\\
	\end{aligned}
	\label{eq29}
\end{equation}
in which $q > 0$, $\varepsilon > 0$, $1 - q\tau > 0$, and the disturbance $\tilde{d}(k) = M^TD_\tau d(k)$. Also, $d_m$ and $d_s$ are the mean and spread of $\tilde{d}(k)$, respectively. Since it is assumed that $d(k)$ is bounded, $\tilde{d}(k)$ will also be bounded. Let the lower and upper bounds on $\tilde{d}(k)$ be $d_l$ and $d_u$, respectively, such that $d_l \leq \tilde{d}(k) \leq d_u$. The mean $d_m$ and the spread $d_s$ of $\tilde{d}(k)$ are related to the lower and upper bounds of $\tilde{d}(k)$ as follows:
\begin{equation}
	\begin{aligned}	
		&d_m = \dfrac{d_l + d_u}{2}, \ \ \ d_s = \dfrac{d_u - d_l}{2}.\\
	\end{aligned}
	\label{eq30}
\end{equation}
A state feedback-based robust discrete-time SMC law that satisfies the reaching law \eqref{eq29} can be computed as
\begin{equation}
	\begin{aligned}	
		w(k) = & -(M^T\Gamma_\tau)^{-1}[(M^T\Phi_\tau - M^T + q\tau M^T)z(k)\\
		& + d_m + (d_s + \varepsilon \tau )\mathrm{sgn}(s(k))],\\
	\end{aligned}
	\label{eq31}
\end{equation}
in which $M^T\Gamma_\tau \neq 0$. The bound on the QSM band, $\xi$, for the state feedback-based discrete-time SMC algorithm can be computed as follows \cite{b19}: $\xi \leq 2d_s + \varepsilon \tau$, with the constraint on the control parameters as $\frac{q\tau ^2 \varepsilon}{2(1 - q\tau)} > d_s$.

\subsection{Stability Analysis}\label{sec4subsec2}
In this section, we propose two unique methods for the sliding surface design of the state feedback-based discrete-time SMC, such that the closed-loop discrete-time model of the magnetic levitation system \eqref{eq26} is stable in the presence of matched uncertainties. We first prove that the closed-loop system \eqref{eq26} is asymptotically stable when the sliding surface and the state trajectories satisfy the ideal QSM conditions. The ideal QSM satisfies $s(k+1) = s(k) = 0$, for $k \geq 0$. Using \eqref{eq26}, $s(k) = M^Tz(k)$, and the above ideal QSM conditions, we obtain
\begin{equation}
	\begin{aligned}	
		s(k+1) &= M^Tz(k+1) = s(k) = M^Tz(k) = 0\\
		&= M^T\Phi_\tau z(k) + M^T\Gamma_\tau w(k) + M^TD_\tau d(k).\\
	\end{aligned}
	\label{eq32}
\end{equation}
Solving for $w(k)$ from \eqref{eq32}, yields
\begin{equation}
	\begin{aligned}	
		w(k) &= - (M^T\Gamma_\tau)^{-1}M^T\Phi_\tau z(k) - d(k),\\
	\end{aligned}
	\label{eq33}
\end{equation}
in which $M^T\Gamma_\tau \neq 0$, and the disturbance $d(k)$ satisfies the matching condition $D_\tau = \Gamma_\tau$. Next, substituting \eqref{eq33} into \eqref{eq26}, the dynamics of the ideal QSM can be written as follows: $z(k+1) = (I - \Gamma_\tau (M^T\Gamma_\tau)^{-1}M^T)\Phi_\tau z(k)$, $M^Tz(k) = 0$. Since the pair $(\Phi_\tau, \Gamma_\tau)$ is controllable, the sliding surface vector, $M^T$, can be designed such that all the eigenvalues of $(I - \Gamma_\tau (M^T\Gamma_\tau)^{-1}M^T)\Phi_\tau $ lie within the unit circle of the origin in the z-domain. As a result, the stability of the closed-loop system \eqref{eq26} for the ideal QSM conditions is guaranteed. Furthermore, since the effect of the matched uncertainty $d(k)$ is canceled in the ideal QSM, the state feedback-based discrete-time SMC is robust to matched uncertainties.

In addition to the above method, we propose an alternative method for designing the sliding surface vector, $M^T$, such that the closed-loop system \eqref{eq26} is stable. In this method, we transform the system \eqref{eq26} to a normal form \cite{b14} similar to the Brunovsky form that we obtained in \eqref{eq8}. Let us define a change of coordinates $z(k) = \Omega \tilde{z}(k)$, such that $\Omega ^{-1}\Gamma_\tau = B = (0 \ \ \ 0 \ \ \ 1)^T$, in which $\Omega \in\mathbb{R}^{3\times3}$ is a nonsingular matrix with rank $\Omega$ = 3. Substituting $z(k) = \Omega \tilde{z}(k)$ into \eqref{eq26}, we obtain the state-space equations in the normal form as follows: $\tilde{z}(k+1) = \tilde{\Phi}_\tau \tilde{z}(k) + B(w(k) + d(k))$, in which $\tilde{z}(k) \in\mathbb{R}^3$ and $\tilde{\Phi}_\tau \in\mathbb{R}^{3\times3}$ are partitioned as
\begin{equation}
	\begin{aligned}	
		&\tilde{z}(k) = \begin{pmatrix} \tilde{z}_1(k)\\ \tilde{z}_2(k)\end{pmatrix}, \ \ \ \tilde{\Phi}_\tau = \begin{pmatrix} \tilde{\Phi}_{11} & \tilde{\Phi}_{12}\\ \tilde{\Phi}_{21} & \tilde{\Phi}_{22}\end{pmatrix}.\\
	\end{aligned}
	\label{eq37}
\end{equation}
In \eqref{eq37}, $\tilde{z}_1(k) \in \mathbb{R}^2$, $\tilde{z}_2(k) \in \mathbb{R}$, $\tilde{\Phi}_\tau = \Omega ^{-1}\Phi_\tau \Omega$, $\tilde{\Phi}_{11} \in\mathbb{R}^{2\times2}$, $\tilde{\Phi}_{12} \in\mathbb{R}^{2\times1}$, $\tilde{\Phi}_{21} \in\mathbb{R}^{1\times2}$, and $\tilde{\Phi}_{22} \in\mathbb{R}^{1\times1}$. The switching surface is also partitioned as $s(k) = M^T z(k) = (M_{11} \ \ M_{12})\tilde{z}(k)$, where $M_{11} \in \mathbb{R}^{1\times2}$ and $M_{12} \in \mathbb{R}^{1\times1}$. Next, using the transformation
\begin{equation}
	\begin{aligned}	
		&\begin{pmatrix} \tilde{z}_1(k)\\ s(k)\end{pmatrix} = L_M\tilde{z}(k) = \begin{pmatrix} I & 0\\ M_{11} & M_{12}\end{pmatrix} \begin{pmatrix} \tilde{z}_1(k)\\ \tilde{z}_2(k)\end{pmatrix},\\
	\end{aligned}
	\label{eq38}
\end{equation}
system $\tilde{z}(k+1) = \tilde{\Phi}_\tau \tilde{z}(k) + B(w(k) + d(k))$ is transformed to
\begin{equation}
	\begin{aligned}	
		\tilde{z}_1(k+1) &= \Psi_{11} \tilde{z}_1(k) + \Psi_{12}s(k),\\
		s(k+1) &= \Psi_{21} \tilde{z}_1(k) + \Psi_{22}s(k) + \tilde{B}(w(k) + d(k)).\\
	\end{aligned}
	\label{eq39}
\end{equation}
In \eqref{eq39}, $\Psi_{11} \in\mathbb{R}^{2\times2}$, $\Psi_{12} \in\mathbb{R}^{2\times1}$, $\Psi_{21} \in\mathbb{R}^{1\times2}$, and $\Psi_{22} \in\mathbb{R}^{1\times1}$ are elements of the matrix $\Psi \in\mathbb{R}^{3\times3}$, and $\tilde{B} = L_MB = (0 \ \ 0 \ \ M_{12})^T$. Using the relation $\Psi = L_M\tilde{\Phi}_\tau L^{-1}_M$, the elements of the matrix $\Psi$ are computed as follows: 
\begin{equation}
	\begin{aligned}	
		\Psi_{11} &=  \tilde{\Phi}_{11} - \tilde{\Phi}_{12}M^{-1}_{12}M_{11}, \ \ \ \Psi_{12} = \tilde{\Phi}_{12}M^{-1}_{12},\\
		\Psi_{21} &= M_{11}\tilde{\Phi}_{11} + M_{12}\tilde{\Phi}_{21} - (M_{11}\tilde{\Phi}_{12} + M_{12}\tilde{\Phi}_{22})M^{-1}_{12}M_{11},\\
		\Psi_{22} &= M_{11}\tilde{\Phi}_{12}M^{-1}_{12} + M_{12}\tilde{\Phi}_{22}M^{-1}_{12}.\\
	\end{aligned}
	\label{eq40}
\end{equation}
Substituting $s(k+1) = s(k) = (M_{11} \ M_{12})\tilde{z}(k)$ = 0 into \eqref{eq39}, the dynamics of the ideal QSM can be written as follows: $\tilde{z}_1(k+1) = (\tilde{\Phi}_{11} - \tilde{\Phi}_{12}M^{-1}_{12}M_{11})\tilde{z}_1(k)$, $\tilde{z}_2(k) = M^{-1}_{12}M_{11}\tilde{z}_1(k)$. Since the pair $(\Phi_\tau, \Gamma_\tau)$ is controllable, the pair $(\tilde{\Phi}_{11}, \tilde{\Phi}_{12})$ is also controllable. Thus, the sliding surface vector, $M^T$, can be designed such that all the eigenvalues of $(\tilde{\Phi}_{11} - \tilde{\Phi}_{12}M^{-1}_{12}M_{11})$ lie within the unit circle of the origin in the z-domain. The above design guarantees the stability of the closed-loop system \eqref{eq26} when the ideal QSM conditions are met.

However, in reality the QSM motion for the state feedback-based discrete-time SMC is not ideal with $\xi \neq 0$. Substituting the discrete-time SMC law \eqref{eq31} into \eqref{eq26}, yields
\begin{equation}
	\begin{aligned}	
		z(k+1) = & \ (I - \Gamma_\tau (M^T\Gamma_\tau)^{-1}M^T)\Phi_\tau z(k) + \Gamma_\tau d(k)\\
		& + (1 - q\tau)\Gamma_\tau(M^T\Gamma_\tau)^{-1}M^Tz(k)\\
		& -\Gamma_\tau(M^T\Gamma_\tau)^{-1}(d_m + (d_s + \varepsilon \tau )\mathrm{sgn}(s(k))).\\
	\end{aligned}
	\label{eq42}
\end{equation}
Equation \eqref{eq42} represents the QSM motion of the closed-loop system \eqref{eq26} within the $2\xi$-band. Smaller the width of the QSM $2\xi$-band, closer will be the system response to that of an equivalent continuous-time SMC feedback law. Due to the finite sampling frequency, the state trajectory moves in a sliding-like QSM manner along the switching surface within a finite $2\xi$-band. As a result of the QSM motion of the state trajectory, the invariance property present in the PI-SMC is lost to some extent in the state feedback-based discrete-time SMC, making it less robust to uncertainties or parametric variations. 

\section{Multirate Output Feedback-Based Discrete-Time Sliding Mode Control}\label{sec5}
The SMC strategies reported in the literature so far for a magnetic levitation system are either based on state feedback or static output feedback. Since all the system states are not directly available for measurement in the magnetic levitation system, separate observers are required to estimate the state variables, thereby increasing the complexity of the system. Also, the observer-based output feedback method has the following drawbacks: a) System states are computed in theoretically infinite time, and b) Error between the computed state and the actual state decreases asymptotically and goes to zero as $t \to \infty$. In the MROF-based discrete-time SMC strategy \cite{b18,b19,b20}, only the multirate-sampled system outputs and the past control inputs are used to compute the control input, overcoming the problem of state estimation. The state feedback-based discrete-time SMC strategy presented earlier is only robust to matched uncertainties. Also, the chattering present in the PI-SMC is significantly reduced in the state feedback-based discrete-time SMC but at the cost of losing invariance. To improve the robustness without compromising the chattering reduction benefits of the state feedback-based discrete-time SMC, mismatched uncertainties like sensor noise and track input disturbance are incorporated in the MROF-based discrete-time SMC design. The MROF-based discrete-time SMC strategy shows better dynamic and steady-state performance in the presence of mismatched uncertainties compared to the state feedback-based discrete-time SMC algorithm. 

\subsection{Multirate Output Feedback Technique}\label{sec5subsec1}
Consider the following $m$-input, $p$-output, $n^{\mathrm{th}}$ order discrete-time system representation with uncertainty in the state equation, sampled at a sampling interval of $\tau$ seconds: $z(k+1) = \Phi_\tau z(k) + \Gamma_\tau w(k) + D_\tau d(k)$, $y(k) = C_\tau z(k)$, in which $z(k)\in\mathbb{R}^n$, $w(k)\in \mathbb{R}^m$, $y(k)\in \mathbb{R}^p$, $d(k)\in \mathbb{R}^q$, $\Phi_\tau \in \mathbb{R}^{n\times n}$, $\Gamma_\tau \in \mathbb{R}^{n\times m}$, $D_\tau \in \mathbb{R}^{n\times q}$, and $C_\tau \in \mathbb{R}^{p\times n}$. For the magnetic levitation system, $n = 3$, $m = 1$, $p = 1$, and $q \leq 3$. The disturbance $d(k)$ is composed of model uncertainties, unknown internal dynamics, and external disturbances. It is assumed here that $d(k)$ is bounded and mismatched with the control input $w(k)$, i.e., the matching condition rank($\Gamma_\tau$) = rank($[\Gamma_\tau \lvert D_\tau ]$) is not necessarily satisfied. The control input $w(k)$ is applied at a sampling interval of $\tau$ s and the sensor output $y(k)$ is measured after every $\rho = \tau/N$ s, where $N$ is an integer greater than or equal to the observability index $O$ of the system. The system sampled at the $\rho$ interval is represented using the quadruplet $(\Phi_\rho, \Gamma_\rho, C_\rho, D_\rho)$, and the system sampled at the $\tau$ interval is represented using the quadruplet $(\Phi_\tau, \Gamma_\tau, C_\tau, D_\tau)$. It is assumed, without loss of generality that the pair $(\Phi_\tau, \Gamma_\tau)$ is controllable, and the pair $(\Phi_\rho, C_\rho)$ is observable. With the assumption that the disturbance $d(k)$ remains unchanged during each $\tau$-interval, the system sampled at an interval of $\rho$ s is as follows: 
\begin{equation}
	\begin{aligned}	
		z(k+1) &= \Phi_\rho z(k) + \Gamma_\rho w(k) + D_\rho d(k),\\ 
		y(k) &= C_\rho z(k),\\
	\end{aligned}
	\label{eq44}
\end{equation}
in which $\Phi_\rho \in \mathbb{R}^{n\times n}$, $\Gamma_\rho \in \mathbb{R}^{n\times m}$, $D_\rho \in \mathbb{R}^{n\times q}$, and $C_\rho \in \mathbb{R}^{p\times n}$. The matrices of the $\tau$-system and the $\rho$-system have the following relationship:
\begin{equation}
	\begin{aligned}	
		&\Phi_\tau = \Phi^N_\rho, \ \ \Gamma_\tau = \Bigg{(}\sum^{N-1}_{i = 0} \Phi^i_\rho \Bigg{)}\Gamma_\rho,\ \ D_\tau = \Bigg{(}\sum^{N-1}_{i = 0}\Phi^i_\rho\Bigg{)}D_{\rho}.\\
	\end{aligned}
	\label{eq45}
\end{equation}
Equation \eqref{eq45} shows that the relationship between the system matrices of the $\tau$-system and the $\rho$-system is independent of $\tau$ or $\rho$ and is only dependent on the value of $N$. Using \eqref{eq44}--\eqref{eq45}, the MROF representation of the system, with the input sampling interval of $\tau$ s and the output sampling interval of $\rho$ s, can be written as follows \cite{b19, b20}:
\begin{equation}
	\begin{aligned}	
		z(k+1) &= \Phi_\tau z(k) + \Gamma_\tau w(k) + D_\tau d(k),\\
		y_{k+1} &= C_0 z(k) + D_0 w(k) + D_y d(k),\\
	\end{aligned}
	\label{eq46}
\end{equation}
in which 
\begin{equation}
	\begin{aligned}	
		y_k &= \begin{pmatrix} y((k-1)\tau)\\ y((k-1)\tau + \rho) \\ y((k-1)\tau + 2\rho)\\ \vdots \\ y(k\tau - \rho) \end{pmatrix}, \ \ \ C_0 = \begin{pmatrix} C_\rho\\ C_\rho\Phi_\rho\\ C_\rho\Phi_\rho^2\\ \vdots \\ C_\rho\Phi_\rho^{N-1} \end{pmatrix},\\
		D_0 &= \begin{pmatrix} 0\\ C_\rho\Gamma_\rho\\ C_\rho(\Phi_\rho \Gamma_\rho + \Gamma_\rho)\\ \vdots \\ C_\rho\sum^{N-2}_{i = 0} \Phi_\rho^i \Gamma_\rho \end{pmatrix}, \ \ \ D_y = \begin{pmatrix} 0\\ C_\rho D_\rho\\ C_\rho(\Phi_\rho D_\rho + D_\rho)\\ \vdots \\ C_\rho\sum^{N-2}_{i = 0} \Phi_\rho^i D_\rho \end{pmatrix}.\\
	\end{aligned}
	\label{eq47}
\end{equation}
In \eqref{eq47}, $y_k$ represents the past $N$ multirate-sampled system outputs. From \eqref{eq46}, the state vector $z(k)$ can be expressed in terms of the output samples $y_{k+1}$, control input $w(k)$, and disturbance $d(k)$ as $z(k) = (C_0^TC_0)^{-1}C_0^T(y_{k+1} - D_0 w(k) - D_y d(k))$. Since the value of $N$ is chosen to be greater than or equal to the observability index of the system, $C_0 \in \mathbb{R}^{pN \times n}$ is a matrix of rank $n$. Moreover, $C_0^TC_0 \in \mathbb{R}^{n \times n}$ is also a matrix of rank $n$, and hence is invertible with $C_0^TC_0 \neq 0$. From \eqref{eq44} and \eqref{eq46}, the state vector $z(k)$ can be represented as a function of the  past control input $w(k-1)$, past multirate output samples $y_k$, and past disturbance $d(k-1)$ as follows:
\begin{equation}
	\begin{aligned}	
		z(k) &= L_w w(k-1) + L_y y_k + L_d d(k-1),\\
	\end{aligned}
	\label{eq49}
\end{equation}
\begin{flushleft} in which  $L_w = \Gamma _\tau - \Phi_\tau (C_0^TC_0)^{-1}C_0^T D_0$, $L_y = \Phi_\tau (C_0^TC_0)^{-1}C_0^T$, $L_d = D_\tau - \Phi_\tau (C_0^TC_0)^{-1}C_0^T D_y$. \end{flushleft}

The MROF-based technique does not make use of the present inputs or outputs, thus providing for a time-delay required for control law implementation. Also, the system states are computed exactly after just one sampling interval as opposed to a theoretically infinite time taken by a Luenberger observer. The MROF-based technique can achieve zero steady-state error once a multirate-sampled output measurement is available. After the first set of output samples are obtained, the MROF-based controller would function exactly like a continuous-time full-state feedback controller.

\subsection{Controller Design}\label{sec5subsec2}
A robust multirate output feedback-based discrete-time SMC algorithm for the magnetic levitation system is derived using the following reaching law:
\begin{equation}
	\begin{aligned}	
		s(k+1) - s(k) = & -q\tau s(k) - \varepsilon \tau \mathrm{sgn}(s(k)) + \tilde{r}(k-1)\\
		&+ \tilde{d}(k) - d_m - r_m - (d_s + r_s)\mathrm{sgn}(s(k)),\\
	\end{aligned}
	\label{eq51}
\end{equation}
in which 
\begin{equation}
	\begin{aligned}	
		&r_l \leq \tilde{r}(k) \leq r_u, \ \ \ r_m = \dfrac{r_l + r_u}{2}, \ \ \ r_s = \dfrac{r_u - r_l}{2},\\
		&\tilde{r}(k-1) = (M^T\Phi_\tau - M^T + q\tau M^T)L_d d(k-1).
	\end{aligned}
	\label{eq52}
\end{equation}
In \eqref{eq51}, $q > 0$, $\varepsilon > 0$, $1 - q\tau > 0$, the disturbance $\tilde{d}(k) = M^TD_\tau d(k)$, and $\tilde{r}(k-1)$ represents the mismatched uncertainties like sensor noise and track input disturbance affecting the system. Also, $r_m$ and $r_s$ are the mean and spread of $\tilde{r}(k)$, respectively, and the switching surface $s(k) = M^Tz(k) = M^TL_w w(k-1) + M^TL_y y_k + M^TL_d d(k-1)$. Now, a robust discrete-time SMC feedback law satisfying the reaching law in \eqref{eq51} can be formulated as follows: $w(k) = -(M^T\Gamma_\tau)^{-1}[d_m + r_m + (d_s + r_s + \varepsilon \tau )\mathrm{sgn}(s(k)) + (M^T\Phi_\tau - M^T + q\tau M^T)z(k) - \tilde{r}(k-1)]$. The above control law has an uncertain component $\tilde{r}(k-1)$. However, by using the state vector $z(k)$ from \eqref{eq49}, the uncertain components in \eqref{eq49} and the control law $w(k)$ cancel out. Thus, the robust multirate output feedback-based discrete-time SMC law can be written as follows:
\begin{equation}
	\begin{aligned}	
		w(k) = & \ F_y y_k + F_w w(k-1) - G_m - G_s\mathrm{sgn}(s(k)),\\
	\end{aligned}
	\label{eq54}
\end{equation}
in which 
\begin{equation}
	\begin{aligned}	
		F_y &= -(M^T\Gamma_\tau)^{-1}(M^T\Phi_\tau - M^T + q\tau M^T)L_y,\\
		F_w &= -(M^T\Gamma_\tau)^{-1}(M^T\Phi_\tau - M^T + q\tau M^T)L_w,\\
		G_m &= (M^T\Gamma_\tau)^{-1}(d_m + r_m),\\
		G_s &= (M^T\Gamma_\tau)^{-1}(d_s + r_s + \varepsilon \tau).		
	\end{aligned}
	\label{eq55}
\end{equation}
The design parameters in \eqref{eq55} can be appropriately tuned to achieve the desired level of robustness, improve the convergence rate, reduce chattering, and decrease the control effort and energy. In \eqref{eq55}, $G_m$ and $G_s$ are the disturbance compensation components of the MROF controller. The components $G_m$ and $G_s$ improve the robustness of the closed-loop system in the presence of mismatched uncertainties like measurement sensor noise and track input disturbance. The disturbance rejection ability of the MROF-based discrete-time SMC is dependent on the appropriate design of the disturbance compensation components.

\begin{figure}[!t]
	\centerline{\includegraphics[height=4.18cm, width=\columnwidth]{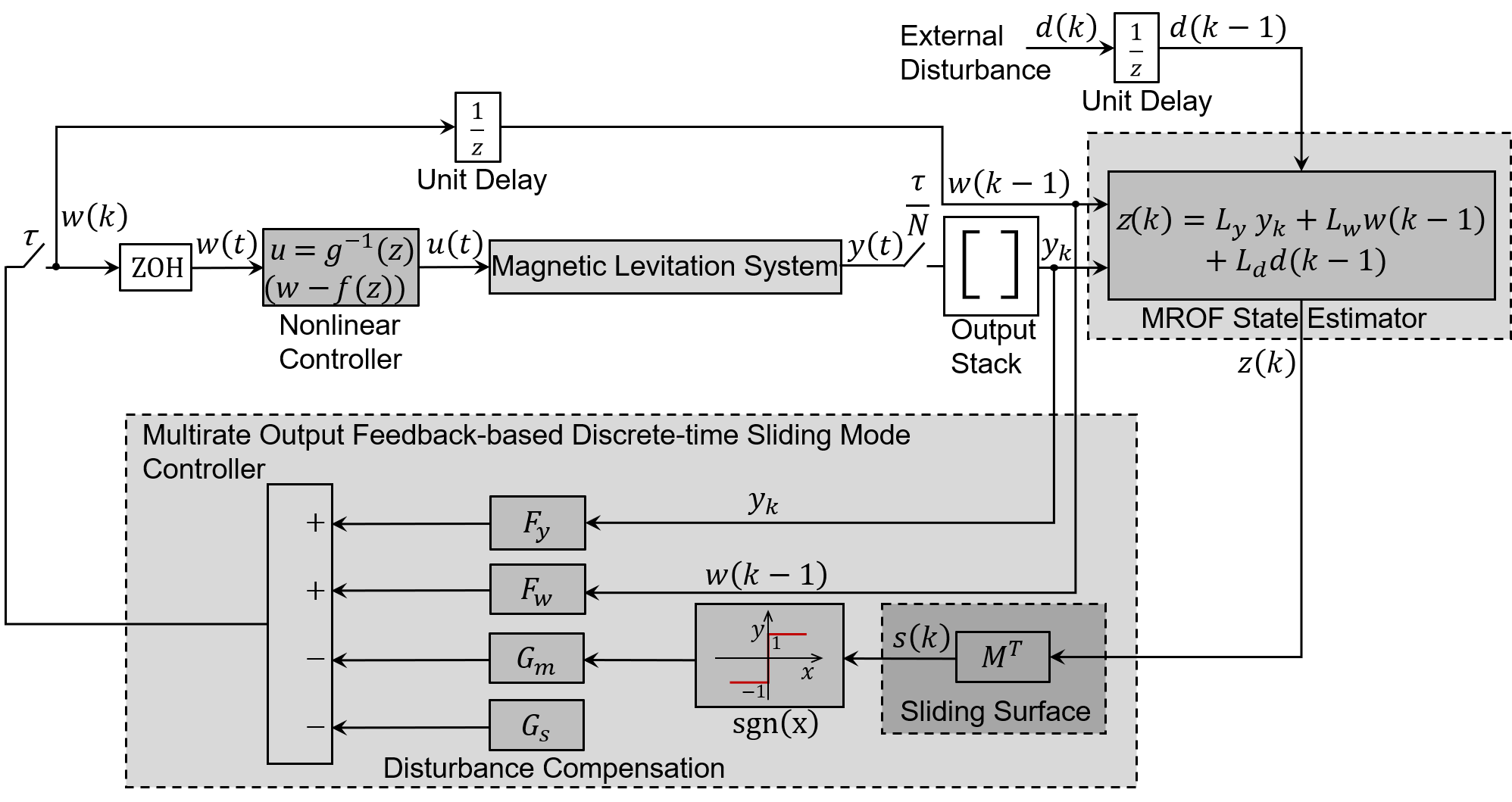}}
	\caption{Schematic of the MROF-based strategy.}
	\label{fig2}
\end{figure}

A schematic of the MROF-based discrete-time SMC applied to a magnetic levitation system is shown in Fig. \ref{fig2}. The overall closed-loop system consists of the following elements: magnetic levitation system nonlinear model, model discretizer, ZOH block, output stack, unit delay blocks, MROF-based state estimator, MROF-based discrete-time SMC, and an outer-loop nonlinear feedback controller. As shown in Fig. \ref{fig2}, the control input $w(k)$ is applied with a sampling interval of $\tau = 0.06$ s, and the output is sampled at a faster rate of $\rho = 0.02$ s. We have chosen $N$ to be equal to the observability index ($O = 3$) of the magnetic levitation system model so that all the matrices in \eqref{eq44} and \eqref{eq46} are square matrices, which further simplifies the calculations. For the MROF representation of the magnetic levitation system \eqref{eq46} with $\tau = 0.06$ s, $\rho = 0.02$ s, and $N = 3$, the system matrices are
\begin{equation}
	\begin{aligned}	
		&\Phi_\tau = \begin{pmatrix} 1 & 0.06 & 0.0018\\ 0 & 1 & 0.06\\ 0 & 0 & 1\end{pmatrix}, \ \ \ \Gamma_\tau = \begin{pmatrix} 0.00\\ 0.0018\\ 0.06\end{pmatrix},\\
		&\Phi_\rho = \begin{pmatrix} 1 & 0.02 & 0.0002\\ 0 & 1 & 0.02\\ 0 & 0 & 1\end{pmatrix}, \ \ \ \Gamma_\rho = \begin{pmatrix} 0.00\\ 0.0002\\ 0.02\end{pmatrix}.\\  
	\end{aligned}
	\label{eq56}
\end{equation}

The switching surface $s(k)$ cannot be exactly computed due to the presence of the uncertainty term $L_d d(k-1)$. However, only
the sign of $s(k)$ is necessary for computation of the control input $w(k)$ in \eqref{eq54}. Since, only the sign is necessary, but not the exact value, $\mathrm{sgn}(s(k))$ is replaced with $\mathrm{sgn}(\tilde{s}(k))$, where $\tilde{s}(k)$ is computed as follows:
\begin{equation}
	\begin{aligned}	
		\tilde{s}(k) &= M^TL_w w(k-1) + M^TL_y y_k + M^Tn_m\\
		&= s(k) - M^Tn(k-1) + M^Tn_m,
	\end{aligned}
	\label{eq58}
\end{equation}
in which $n_l \leq L_d d(k-1) = n(k-1) \leq n_u$, and
\begin{equation}
	\begin{aligned}	
		&n_m = \dfrac{n_l + n_u}{2}, \ \ \ n_s = \dfrac{n_u - n_l}{2}.\\
	\end{aligned}
	\label{eq59}
\end{equation}
In \eqref{eq59}, $n_m$ and $n_s$ are the mean and spread of $n(k-1)$, respectively. When $\lvert \tilde{s}(k)\rvert > n_s$, the value of $\mathrm{sgn}(s(k)) = \mathrm{sgn}(\tilde{s}(k))$. However, when $\lvert \tilde{s}(k)\rvert \leq n_s$, the sign of $s(k)$ cannot be determined accurately as $s(k)$ is in the range of $\tilde{s}(k) \pm n_s$. Hence, in order to ensure QSM motion of the state trajectory, the width of the QSM band should be such that it encompasses this ambiguous band of $\lvert s(k)\rvert \leq 2n_s$. The bound on the QSM band, $\xi$, for the MROF-based discrete-time SMC algorithm is computed as follows \cite{b19}: 
\begin{equation}
	\begin{aligned}	
		&\xi < \dfrac{2(d_s + r_s) + n_s + \varepsilon \tau}{(1 - q\tau)},\\
	\end{aligned}
	\label{eq60}
\end{equation}
with the constraints on the control parameters as $\frac{q\tau ^2 \varepsilon}{2(1 - q\tau)} > (d_s + r_s)$, and $2(d_s + r_s) + \varepsilon \tau > 2n_s$. From \eqref{eq60}, we can see that the MROF-based discrete-time SMC method has an increased bound on the QSM band compared to the state feedback-based discrete-time SMC algorithm.

\subsection{Stability Analysis}\label{sec5subsec3}
In this section, we prove that the MROF-based discrete-time sliding mode controller asymptotically stabilizes the nonlinear model \eqref{eq2} of the magnetic levitation system.

\begin{theorem}\label{thm2} Consider the following MROF-based discrete-time SMC law for the magnetic levitation system:
	\begin{equation}
		\begin{aligned}	
			w(k) = & \ F_y y_k + F_w w(k-1) - G_m - G_s\mathrm{sgn}(\tilde{s}(k))\\
			\approxeq & -(M^T\Gamma_\tau)^{-1}[p_m + p_s \mathrm{sgn}(\tilde{s}(k))\\
			& + M^T(\Phi_\tau - I)(z(k) - n(k-1) + n_m)],\\
		\end{aligned}
		\label{eq61}
	\end{equation}
	in which $F_y$, $F_w$, $G_m$, and $G_s$ are as given in \eqref{eq55}; $p_m = d_m + r_m$, $p_s = d_s + r_s + \varepsilon \tau $, and $n(k-1) = L_d d(k-1)$; $d_m$ and $d_s$ are obtained from \eqref{eq30}; $r_m$ and $r_s$ are as given in \eqref{eq52}; and $\tilde{s}(k)$ and $n_m$ are obtained from \eqref{eq58} and \eqref{eq59}, respectively. Let us assume that the control law \eqref{eq61} satisfies the following conditions:
	\begin{align}	
		&\tilde{z}(k) = z(k) - n(k-1) + n_m,
		\label{eq62}\\
		&\frac{q\tau ^2 \varepsilon}{2(1 - q\tau)} > (d_s + r_s), \ \ \ 2(d_s + r_s) + \varepsilon \tau > 2n_s,
		\label{eq63}\\
		&\theta_l \leq M^T \Phi_\tau n(k) = \theta(k) \leq \theta_u, \ \ \ n(k) = L_d d(k),
		\label{eq64}\\
		&\theta_m = \dfrac{\theta_l + \theta_u}{2} = M^T \Phi_\tau n_m, \ \ \ \theta_s = \dfrac{\theta_u - \theta_l}{2},
		\label{eq65}\\
		&\gamma_l \leq M^T n(k) = \gamma(k) \leq \gamma_u, \ \ \ \tilde{d}(k) = M^TD_\tau d(k),
		\label{eq66}\\
		&\gamma_m = \dfrac{\gamma_l + \gamma_u}{2} = M^T n_m, \ \ \ \gamma_s = \dfrac{\gamma_u - \gamma_l}{2},
		\label{eq67}\\
		&p_s \geq \theta_s + \gamma_s + d_s + \sigma_s,\ \ \ \sigma_s > r_m > 0,
		\label{eq68}\\	
		&p_s + r_m - (\theta_s + \gamma_s + d_s) < 2\xi.
		\label{eq69}
	\end{align}
	The MROF-based control law \eqref{eq61} satisfying the conditions \eqref{eq62}--\eqref{eq69} would achieve quasi-sliding mode for the closed-loop discrete-time system \eqref{eq46} in a finite time. That is, $\forall z(0) \in \mathbb{R}^3$, $\exists k^*$ such that $z(k) \in \tilde{\Theta} \subset \Theta$, $\forall k > k^*$, where $k^* > 0$ and the QSM band $\Theta = \{z(k) \lvert -\xi \leq s(k) \leq \xi\}$. The QSM motion of the state trajectories would guarantee the asymptotic convergence of the system states to their desired equilibrium values, i.e., $\lim\limits_{t \to \infty} \lVert x(t) - x_d(t) \rVert = 0$.
\end{theorem}

\textit{Proof:} Consider the Lyapunov function $V(k) = \tilde{s}^2(k)$. The sliding surface $\lvert \tilde{s}(k)\rvert$ decreases monotonically and the system achieves QSM motion in a finite time if the increment $\Delta V(k) = V(k+1) - V(k) < 0$ for $\tilde{s}(k) \neq 0$. Using $V(k) = \tilde{s}^2(k)$, the increment of $V(k)$ can be written as
\begin{equation}
	\begin{aligned}	
		\Delta V(k) &= V(k+1) - V(k) = \tilde{s}^2(k+1) - \tilde{s}^2(k)\\
		&= (\tilde{s}(k+1) + \tilde{s}(k))(\tilde{s}(k+1) - \tilde{s}(k)).
	\end{aligned}
	\label{eq70}
\end{equation}
Substituting $\Delta \tilde{s}(k) = \tilde{s}(k+1) - \tilde{s}(k)$ into \eqref{eq70}, we get
\begin{equation}
	\begin{aligned}	
		\Delta V(k) &= [\Delta \tilde{s}(k) + 2\tilde{s}(k)]\Delta \tilde{s}(k).
	\end{aligned}
	\label{eq71}
\end{equation}
Using \eqref{eq62}, $\Delta \tilde{s}(k)$ can be written as follows:
\begin{equation}
	\begin{aligned}	
		\Delta \tilde{s}(k) =& \ \tilde{s}(k+1) - \tilde{s}(k) = M^T(\tilde{z}(k+1) - \tilde{z}(k))\\
		=& \ M^T(z(k+1) - n(k) + n_m - z(k)\\ 
		& + n(k-1) - n_m)\\
		=& \ M^T(\Phi_\tau z(k) + \Gamma_\tau w(k) + D_\tau d(k) - z(k))\\
		& + M^T(n(k-1) - n(k)).\\
	\end{aligned}
	\label{eq72}
\end{equation}
Substituting the control law \eqref{eq61} into \eqref{eq72}, yields
\begin{equation}
	\begin{aligned}	
		\Delta \tilde{s}(k) =& \ M^T(\Phi_\tau - I)z(k) - p_m - p_s \mathrm{sgn}(\tilde{s}(k))\\
		& - M^T(\Phi_\tau - I)(z(k) - n(k-1) + n_m)\\
		& + M^TD_\tau d(k) + M^T(n(k-1) - n(k))\\
		=& \ (M^T\Phi_\tau n(k-1) - M^T\Phi_\tau n_m) - p_s \mathrm{sgn}(\tilde{s}(k))\\
		& + (M^Tn_m - M^Tn(k)) + (M^TD_\tau d(k) - p_m).\\
	\end{aligned}
	\label{eq73}
\end{equation}
Substituting the conditions in \eqref{eq64}--\eqref{eq67} into \eqref{eq73}, we get
\begin{equation}
	\begin{aligned}	
		\Delta \tilde{s}(k) =& \ (\theta(k-1) - \theta_m) + (\gamma_m - \gamma(k))\\
		& + (\tilde{d}(k) - d_m) - p_s \mathrm{sgn}(\tilde{s}(k)) - r_m.\\		
	\end{aligned}
	\label{eq74}
\end{equation}

First, we consider the case of $\tilde{s}(k) > \xi$. For $\tilde{s}(k) > \xi$, using \eqref{eq65} and \eqref{eq67}, \eqref{eq74} can be written as
\begin{equation}
	\begin{aligned}	
		\Delta \tilde{s}(k) \leq & \ \bigg{(}\theta_u - \dfrac{(\theta_l + \theta_u)}{2}\bigg{)} + \bigg{(}\dfrac{(\gamma_l + \gamma_u)}{2} - \gamma_l\bigg{)} \\
		& + \bigg{(}d_u - \dfrac{(d_l + d_u)}{2}\bigg{)} - p_s - r_m\\	
		\leq & \ \theta_s + \gamma_s + d_s - p_s - r_m.\\		
	\end{aligned}
	\label{eq75}
\end{equation}
Substituting the relations in \eqref{eq68} into \eqref{eq75}, yields
\begin{equation}
	\begin{aligned}	
		\Delta \tilde{s}(k) \leq & \ -\sigma_s - r_m < 0.\\	
	\end{aligned}
	\label{eq76}
\end{equation}
Furthermore, for $\tilde{s}(k) > \xi$, using \eqref{eq69} and \eqref{eq75}, we obtain
\begin{equation}
	\begin{aligned}	
		[\Delta \tilde{s}(k) + 2\tilde{s}(k)] \geq & \ \theta_s + \gamma_s + d_s - p_s - r_m + 2\xi\\	
		> & \ - 2\xi + 2\xi > 0.\\	
	\end{aligned}
	\label{eq77}
\end{equation}
From \eqref{eq76} and \eqref{eq77}, $\forall \tilde{s}(k) > \xi$, $\Delta \tilde{s}(k) < 0$ and $[\Delta \tilde{s}(k) + 2\tilde{s}(k)] > 0$. From \eqref{eq71}, we can conclude that the Lyapunov function increment $\Delta V(k) < 0$ for all $\tilde{s}(k) > \xi$. This proves \eqref{eq28a}, i.e., $\tilde{s}(k) > \xi \Rightarrow 0 < \tilde{s}(k+1) < \tilde{s}(k)$.

Next, we consider the case of $\tilde{s}(k) < -\xi$. For $\tilde{s}(k) < -\xi$, using the relations in \eqref{eq65} and \eqref{eq67}, \eqref{eq74} can be written as
\begin{equation}
	\begin{aligned}	
		\Delta \tilde{s}(k) \geq & \ \bigg{(}\theta_l - \dfrac{(\theta_l + \theta_u)}{2}\bigg{)} + \bigg{(}\dfrac{(\gamma_l + \gamma_u)}{2} - \gamma_u\bigg{)} \\
		& + \bigg{(}d_l - \dfrac{(d_l + d_u)}{2}\bigg{)} + p_s - r_m\\	
		\geq & \ -\theta_s - \gamma_s - d_s + p_s - r_m.\\		
	\end{aligned}
	\label{eq78}
\end{equation}
Substituting the relations in \eqref{eq68} into \eqref{eq78}, yields
\begin{equation}
	\begin{aligned}	
		\Delta \tilde{s}(k) \geq & \ \sigma_s - r_m > 0.\\	
	\end{aligned}
	\label{eq79}
\end{equation}
Also, for $\tilde{s}(k) < -\xi$, using \eqref{eq69} and \eqref{eq78}, we obtain
\begin{equation}
	\begin{aligned}	
		[\Delta \tilde{s}(k) + 2\tilde{s}(k)] \leq & \ - \theta_s - \gamma_s - d_s + p_s - r_m - 2\xi\\	
		< & \ 2\xi - 2r_m - 2\xi < - 2r_m < 0.\\	
	\end{aligned}
	\label{eq80}
\end{equation}
From \eqref{eq79} and \eqref{eq80}, $\forall \tilde{s}(k) < -\xi$, $\Delta \tilde{s}(k) > 0$ and $[\Delta \tilde{s}(k) + 2\tilde{s}(k)] < 0$. From \eqref{eq71}, we can conclude that the Lyapunov function increment $\Delta V(k) < 0$ for all $\tilde{s}(k) < -\xi$. This proves \eqref{eq28b}, i.e., $\tilde{s}(k) < -\xi \Rightarrow 0 > \tilde{s}(k+1) > \tilde{s}(k)$.

Finally, we consider the case of $\lvert \tilde{s}(k)\rvert \leq \xi$. For $0 \leq \tilde{s}(k) \leq \xi$, using the relations in \eqref{eq65}, \eqref{eq67}, and \eqref{eq68}, \eqref{eq74} can be written as
\begin{equation}
	\begin{aligned}	
		\tilde{s}(k+1) \leq & \ \tilde{s}(k) + \theta_s + \gamma_s + d_s - p_s - r_m\\	
		\leq & \ \xi - \sigma_s - r_m < \xi.
	\end{aligned}
	\label{eq81}
\end{equation}
Similarly, for $-\xi \leq \tilde{s}(k) \leq 0$, 
using the relations in \eqref{eq65}, \eqref{eq67}, and \eqref{eq68}, \eqref{eq74} can be written as
\begin{equation}
	\begin{aligned}	
		\tilde{s}(k+1) \geq & \ \tilde{s}(k) - \theta_s - \gamma_s - d_s + p_s - r_m\\	
		\geq & \ - \xi + \sigma_s - r_m > - \xi.
	\end{aligned}
	\label{eq82}
\end{equation}
From \eqref{eq81} and \eqref{eq82}, we have proved \eqref{eq28c}, i.e., $\lvert s(k)\rvert \leq \xi \Rightarrow \lvert s(k+1)\rvert \leq \xi$.

Thus, we have proved the conditions \eqref{eq28a}--\eqref{eq28c} necessary to achieve QSM motion for the closed-loop discrete-time system \eqref{eq46}. Therefore, we can conclude that the QSM band $\Theta$ is attractive and positively invariant. That is, $\forall z(k) \notin \Theta$, $\exists k^*$ such that $\forall k > k^*$, $z(k) \in \Theta$; and $\forall z(0) \in \mathbb{R}^3$, $\exists k^*$ such that $\forall k > k^*$, $z(k) \in \tilde{\Theta} \subset \Theta$, in which $\tilde{\Theta} = \{z(k) \lvert - \xi + \sigma_s - r_m \leq s(k) \leq \xi - \sigma_s - r_m\}$. The stability of the closed-loop discrete-time system \eqref{eq46} implies that the closed-loop discrete-time nonlinear model \eqref{eq26a} is also asymptotically stable. We know that the system stability is conserved on discretization for a sliding mode controller based on a linear sliding hyperplane. Thus, the stability of the discrete-time nonlinear model \eqref{eq26a} implies that the continuous-time nonlinear model \eqref{eq5} is also asymptotically stable. Also, we proved in Section \ref{sec3subsec2} that a state co-ordinate transformation along with feedback linearization does not affect the stability of the overall closed-loop nonlinear system \eqref{eq2}. Thus, we can conclude that the MROF-based discrete-time sliding mode controller asymptotically stabilizes the overall nonlinear model \eqref{eq2} of the magnetic levitation system in the original coordinates and the system states converge to their desired equilibrium values, i.e., $\lim\limits_{t \to \infty} \lVert x(t) - x_d(t) \rVert = 0$.

\section{Simulation and Comparison}\label{sec6}

In this section, the efficacy of the proposed controllers is verified by simulating the magnetic levitation system in MATLAB with the simulation parameters as given in Table \ref{tab1}. The real-time implementation of the proposed sliding mode controllers is possible by using the Embedded Coder function in MATLAB as shown in Fig. \ref{fig1}. To provide a comparative analysis of the proposed control schemes, the following four performance indicators are utilized: a) Integral of the absolute value of the error (IAE), b) Integral of the time multiplied by the absolute value of the error (ITAE), c) Settling time ($t_s$), and d) Maximum deviation of the control input from the steady-state value ($e_{\Delta\mathrm{max}}$). The formulas for calculating IAE, ITAE, and $e_{\Delta\mathrm{max}}$ are as follows:
\begin{equation}
	\begin{aligned}	
		& IAE = \int_0^{t_f}\lvert x_{1d} - x_1\rvert \mathrm{d}t,\ \ e_{\Delta\mathrm{max}} = \lvert e_{\mathrm{max}} - e_\mathrm{ss}\rvert,\\
		& ITAE = \int_0^{t_f}t\lvert x_{1d} - x_1\rvert\mathrm{d}t,\\
	\end{aligned}
	\label{eq83}
\end{equation}
in which $t_f$ is the total running time, $e_{\mathrm{max}}$ is the maximum value of the control input, and $e_\mathrm{ss}$ is the steady state value of the control input.
\begin{figure}[!t]
	\centerline{\includegraphics[height = 7.8cm, width=\columnwidth]{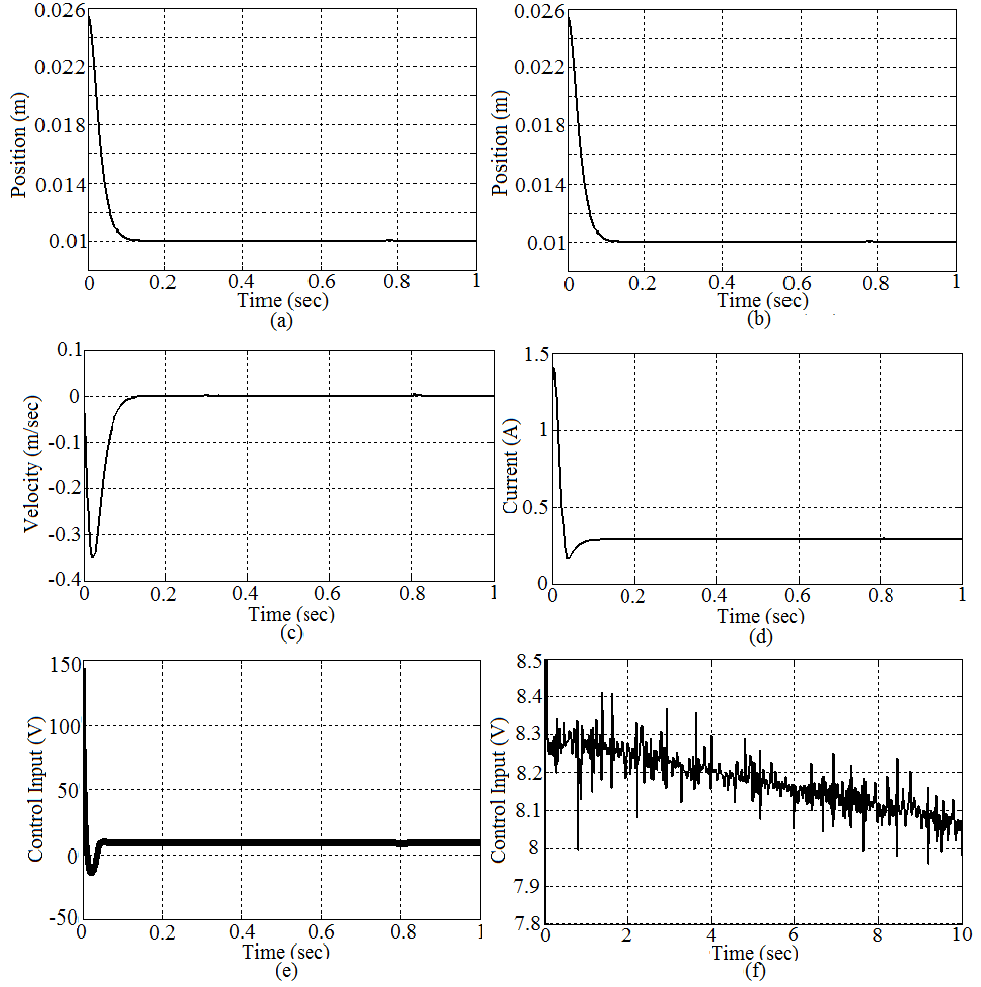}}
	\centerline{\includegraphics[height = 2.6cm, width=\columnwidth]{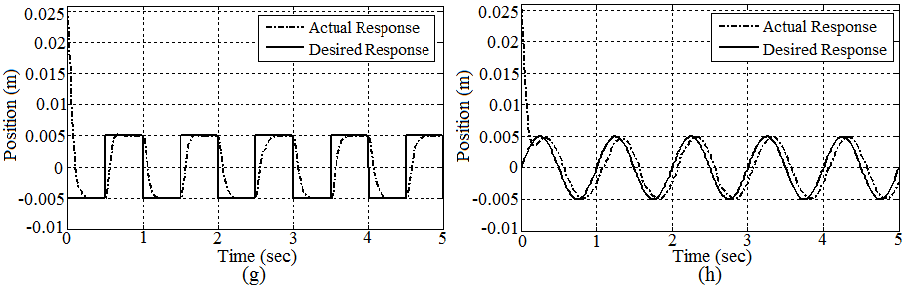}}
	\caption{Steady-state and tracking response plots for the PI-SMC. (a) Position for $m$ = 11.87 g, (b) position for $m$ = (11.87 + 30$\%$) g, (c) velocity, (d) current, (e) control input, (f) enlarged control input, (g) square wave, and (h) sine wave tracking response.}
	\label{fig3}
\end{figure}
\begin{figure}[!t]
	\centerline{\includegraphics[height = 5.2cm, width=\columnwidth]{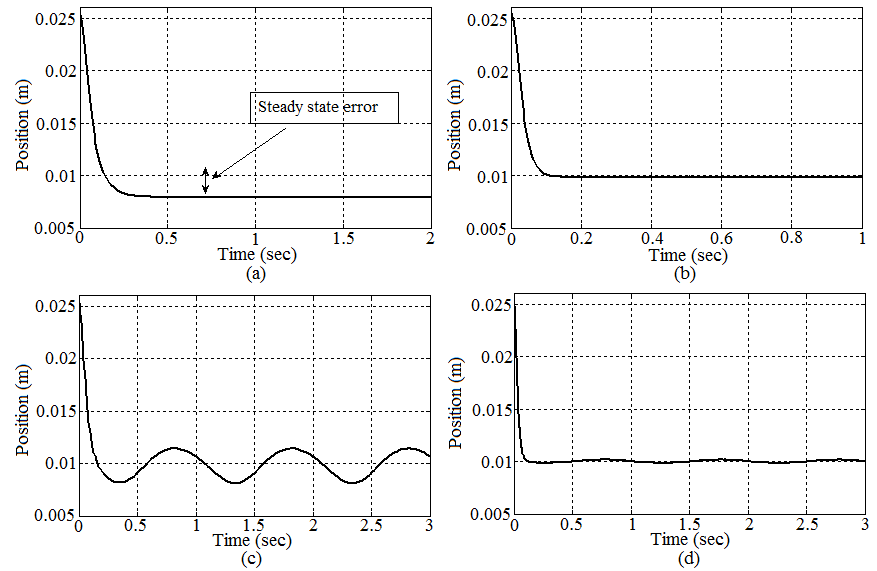}}
	\caption{Disturbance analysis position response plots for the feedback linearization controller \cite{b2} and the PI-SMC. (a) Feedback linearization controller with $d_i(z,t) = 1$, (b) PI-SMC with $d_i(z,t) = 1$, (c) feedback linearization controller with $d_i(z,t) = \sin(2\pi t)$, and (d) PI-SMC with $d_i(z,t) = \sin(2\pi t)$.}
	\label{fig4}
\end{figure}

The PI-SMC design parameters are chosen as follows: $k_4 = 0.1$, $k_5 = 5$, $k_0 = 6$, $\alpha = 0.5$, and the poles of the matrix $A + BK$ are located at $-30$, $-40$, and $-50$. Let $M^T = (1200 \ \ 70 \ \ 1)$ so that the poles of $M$ are located at $-30$ and $-40$. Figs. \ref{fig3}(a) and (b) show position response plots for the PI-SMC for a nominal mass of the ferromagnetic ball and when the mass value is changed by +30$\%$, respectively. The ball position converges to its steady-state value of 0.01 m even when the ball mass is changed by +30$\%$. Thus, we can conclude that the PI-SMC is robust to changes in the ball mass. The position attains its desired steady-state value of 0.01 m in approximately 0.15 s. Figs. \ref{fig3}(c) and (d) show velocity and current response plots for the PI-SMC, respectively. The velocity attains its steady-state value of 0 m/s, and the current converges to its desired steady-state value of 0.2884 A. Figs. \ref{fig3}(e) and (f) show control input response plots for the PI-SMC between 0--1 s and 0--10 s, respectively. The control input in Fig. \ref{fig3}(f) shows that the input voltage fluctuates rapidly between 8--8.4 V. Fig. \ref{fig3}(g) shows the position response plot for the PI-SMC in which the desired signal is a square wave of magnitude 0.005 m and frequency 1 Hz. Fig. \ref{fig3}(h) shows the position response plot for the PI-SMC in which the desired signal is a sine wave with $x_{1d} = 0.005\sin(2\pi t)$. The actual response in both square and sinusoidal wave tracking is close to the desired response.

\begin{table}
	\setlength{\tabcolsep}{6pt}
	\small
	\centering
	\caption{Settling time for different SMC methods}
	\begin{tabular}{|c|c|}
		\hline
		Controller& 
		Settling time $t_s$ (s)\\ \hline
		PI-SMC&
		0.15\\ \hline
		Improved DPRL-I-SMC \cite{b8}&
		1.1\\ \hline
		FOSMC \cite{b9}&
		2.1\\ \hline
		AFTSMC \cite{b7}&
		7\\ \hline
	\end{tabular}
	\label{tab2}
\end{table}

Table \ref{tab2} provides numerical values of the settling time $t_s$ for the PI-SMC, improved DPRL-I-SMC \cite{b8}, FOSMC \cite{b9}, and AFTSMC \cite{b7} control schemes. The proposed PI-SMC scheme has a considerably lower settling time than the improved DPRL-I-SMC, FOSMC, and AFTSMC control schemes reported in the literature. In summary, the proposed PI-SMC scheme outperforms the other three schemes in terms of the convergence rate of the state trajectory to the sliding surface and finally to the equilibrium point.

In this section, the robustness of the PI-SMC is validated. The disturbance $d(z, t)$ is bounded and satisfies the conditions given in Assumption \ref{assumption1}. We consider both the matched $d_3(z, t)$ and the mismatched uncertainties, $d_1(z, t)$ and $d_2(z, t)$, which exist in a practical magnetic levitation system. The mismatched uncertainties are assumed to be zero in the sinusoidal disturbance simulation studies reported in \cite{b7, b8}. Figs. \ref{fig4}(a) and (b) show position response plots for a feedback linearization controller \cite{b2} and the PI-SMC, respectively, in the presence of a disturbance of constant amplitude $d_i(z,t) = 1$ for $i$ = 1, 2, 3. For the feedback linearization controller, we see a small steady-state error in which the ball settles to a new value of 0.0079 m. For the PI-SMC, the ball settles to its desired steady-state value of 0.01 m with an infinitely small steady-state error. Figs. \ref{fig4}(c) and (d) show position response plots for the feedback linearization controller and the PI-SMC, respectively, in the presence of a sinusoidal disturbance $d_i(z,t) = \sin(2\pi t)$ for $i$ = 1, 2, 3. These results show that the ball oscillations about the reference position of 0.01 m are significantly lower for the PI-SMC than the feedback linearization controller. The simulation results show that the PI-SMC is more robust than the feedback linearization controller in the presence of mismatched uncertainties. Also, the closed-loop stability of the PI-SMC in the presence of mismatched uncertainties, which we proved in Theorem \ref{thm1}, is verified via simulation.

\begin{figure}[!t]
	\centerline{\includegraphics[height=3cm, width=4.4cm]{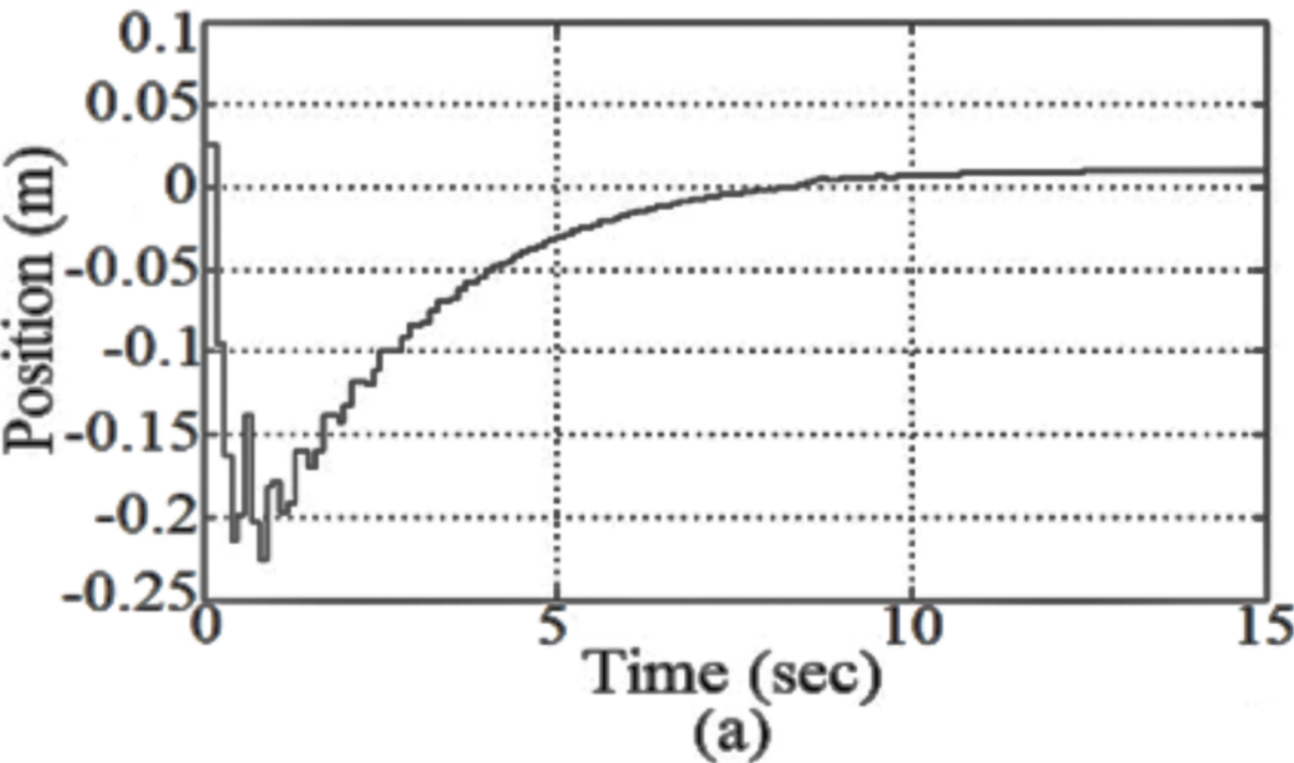}}
	\centerline{\includegraphics[width=\columnwidth]{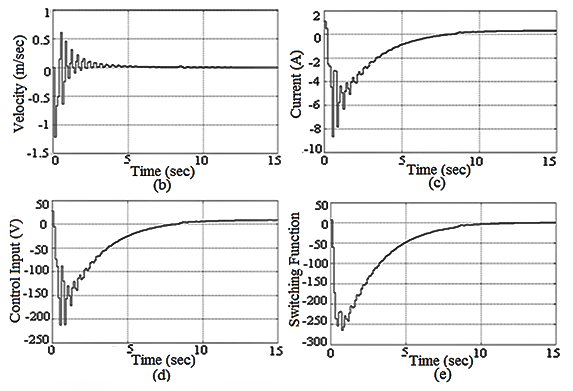}}
	\caption{System response plots for the state feedback-based discrete-time SMC. (a) Position, (b) velocity, (c) current, (d) control input, and (e) switching surface response plots.}
	\label{fig5}
\end{figure}

The state feedback-based discrete-time SMC design parameters are chosen as follows: $q = 0.4$, $\varepsilon = 0.3$, $d_l = -0.001$, $d_u = 0.005$, $d_m = 0.002$, $d_s = 0.003$, and $\tau = 0.1$ s. Let $M^T = (60000 \ \ 4700 \ \ 120)$ so that the eigenvalues of $(I - \Gamma_\tau (M^T\Gamma_\tau)^{-1}M^T)\Phi_\tau $ as well as $(\tilde{\Phi}_{11} - \tilde{\Phi}_{12}M^{-1}_{12}M_{11})$ lie within the unit circle of the origin in the z-domain. These parameters are tuned to obtain the best system response with minimum control chattering. The bound on the QSM band $\xi \leq 2d_s + \varepsilon \tau = 0.036$. Figs. \ref{fig5} (a), (b), (c), (d), and (e) show position $x_1$, velocity $x_2$, current $x_3$, control input $u$, and switching function $s(k)$ response plots for a state feedback-based discrete-time SMC, respectively. The ball attains its steady-state position of 0.01 m in approximately 14 s. The steady-state value of the current is 0.289 A, which is within $\pm$ 0.006 A of the desired steady-state value of 0.2884 A, and the steady-state value of the control input $u$ is 8.28 V. Although it appears that the state variables attain their respective steady-state values, closer examination of the plots reveal that the states oscillate slightly about their desired values. The current fluctuates between 0.2865--0.2915 A, and the velocity of the ball varies between $\pm$ 0.001 m/s. This variation is explained by the QSM motion of the state trajectory. It can be seen from Fig. \ref{fig5}(e) that the switching surface becomes approximately 0, i.e., it enters the QSM band in 12 to 14 seconds.

\begin{figure}[!t]
	\centerline{\includegraphics[width=\columnwidth]{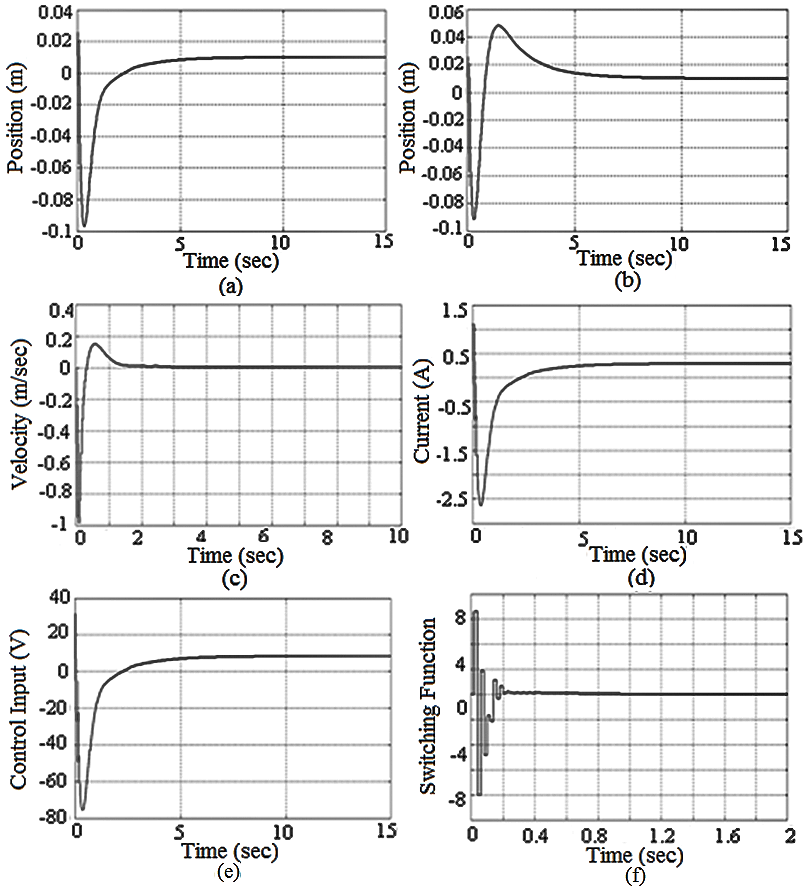}}
	\caption{System response plots for the MROF-based discrete-time SMC. (a) Position for $q = 3$, (b) position for $q = 2$, (c) velocity, (d) current, (e) control input, and (f) switching surface plots for $q = 3$.}
	\label{fig6}
\end{figure}

The MROF-based discrete-time SMC design parameters are chosen as follows: $q = 3$, $\varepsilon = 1$, $d_l = -0.008$, $d_u = 0.014$, $d_m = 0.003$, $d_s = 0.011$, $r_l = -0.002$, $r_u = 0.013$, $r_m = 0.0055$, $r_s = 0.0075$, $n_l = -0.009$, $n_u = 0.015$, $n_m = 0.003$, $n_s = 0.012$, $N = 3$, $\tau = 0.06$ s, and $\rho = 0.02$ s. The sliding surface vector, $M^T = (0.66 \ \ 1 \ \ 0.12)$, is chosen such that the QSM motion has the desired dynamics and the closed-loop system \eqref{eq46} is stable. The design parameters are tuned to obtain the best system response. The bound on the QSM band for the MROF-based controller is calculated using \eqref{eq60} to get $\xi <  0.133$, which is around $3.69$ times the QSM bound for the state feedback-based discrete-time SMC. Figs. \ref{fig6}(a) and (b) show position response plots for the controller gain $q = 3$ and $q = 2$, respectively. For $q = 3$, the ball attains its desired steady-state position of 0.01 m in 8 s, where as for $q = 2$, the ball attains its desired steady-state position in 11 s. Also, the peak overshoot for $q = 2$ is more than that for $q = 3$. Thus, $q = 3$ gives the best steady-state and transient response for all possible values of $q$. Figs. \ref{fig6}(c), (d), (e), and (f) show velocity $x_2$, current $x_3$, control input $u$, and switching function $\tilde{s}(k)$ response plots for the MROF-based discrete-time SMC, respectively. The steady-state value of the current is 0.2883 A, which is within a $\pm$ 0.0001 A variation from the desired steady-state value of 0.2884 A, and the steady-state value of the control input is 8.2722 V.
\begin{figure}[!t]
	\centerline{\includegraphics[width=\columnwidth]{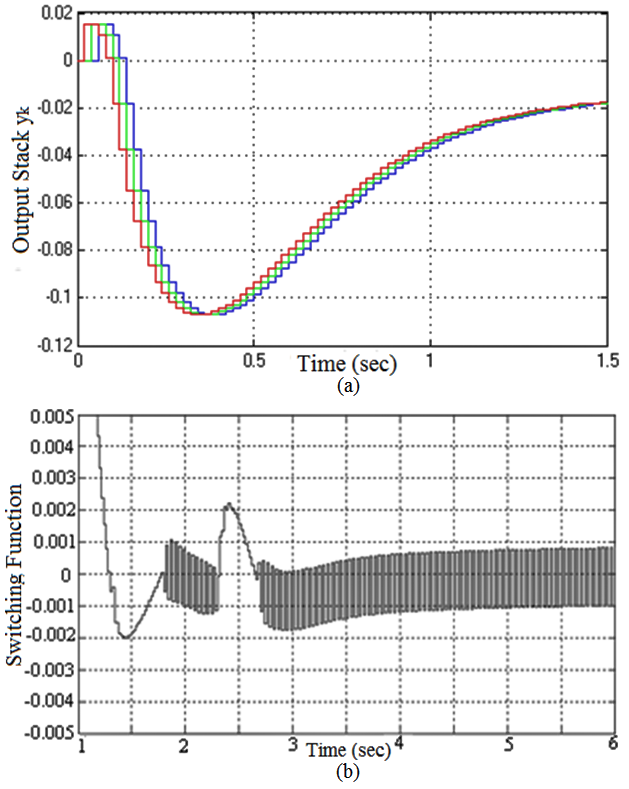}}
	\caption{(a) Output stack, $y_k$, and (b) magnified switching surface response plots for the MROF-based discrete-time SMC.}
	\label{fig7}
\end{figure}
\begin{figure}[!t]
	\centerline{\includegraphics[width=\columnwidth]{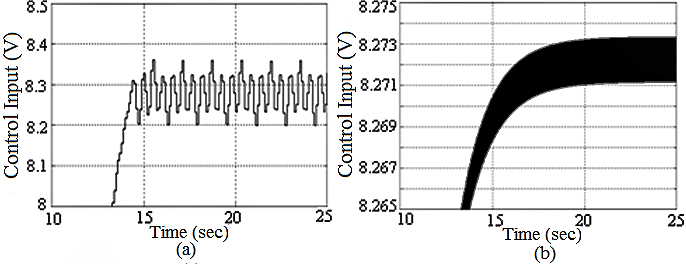}}
	\caption{Chattering observed in (a) the state feedback-based discrete-time SMC and (b) the MROF-based discrete-time SMC.}
	\label{fig8}
\end{figure} 

Fig. \ref{fig7}(a) shows the output stack, $y_k$, which is the past three multirate-sampled system outputs. There is a delay of $\tau$ = 0.02 s between the consecutive (red-green-blue) multirate-sampled system outputs. The magnified plot of the switching surface $\tilde{s}(k)$ in Fig. \ref{fig7}(b) shows that it reaches the QSM band within 2--3 s and, thereafter it remains within the QSM band. This time of 2--3 s is significantly less compared to the 12--14 s time taken by the state feedback-based discrete-time SMC to reach the QSM band. The actual width of the QSM band is approximately 0.0018, which is much lower than the bound on the QSM band $\xi <  0.133$. This is because the bound is calculated for the worst-case scenario of the mismatched uncertainty which may rarely persist for a long time in the magnetic levitation system.
\begin{table}
	\setlength{\tabcolsep}{2pt}
	\small
	\centering
	\caption{Quantitative performance comparison of the proposed controllers}
	\begin{tabular}{|c|c|c|c|}
		\hline
		Criteria& 
		PI-SMC&
		DSMC& 		 
		MROF-DSMC\\ \hline
		$t_s$ (s)&
		0.15&
		14&
		8\\ \hline
		IAE&
		$8.8 \times 10^{-4}$&
		0.805&
		$1.05 \times 10^{-2}$\\ \hline
		ITAE&
		$1.0648 \times 10^{-5}$&
		9.8612&
		$1.641 \times 10^{-2}$\\ \hline
		$e_{\Delta\mathrm{max}}$ (V)&
		136.72&
		218.28&
		83.2722\\ \hline
		Chattering $\Delta e$ (V)&
		0.4&
		0.16&
		0.0022\\ \hline
	\end{tabular}
	\label{tab3}
\end{table}

\begin{table}
	\setlength{\tabcolsep}{1.5pt}
	\small
	\centering
	\caption{Qualitative performance comparison of the proposed controllers}
	\begin{tabular}{|c|c|c|c|c|}
		\hline
		Controller & Robustness & Steady & Chattering & Convergence\\
		&  & State Error &  & Rate\\ \hline
		PI-SMC & Excellent & None & High & Fastest\\ \hline
		DSMC & Fair & Exists & Low & Slow\\ \hline
		MROF-DSMC & Superior & Negligible & Low/ & Medium/\\
		          &  &  & Medium & Fast\\ \hline
	    Improved & Good & Negligible & Low & Faster\\
		DPRL-I-SMC \cite{b8} &  &  &  & \\ \hline
		FOSMC \cite{b9} & Fair & Exists & Medium/ & Fast\\
		                &  &  & High & \\ \hline
		AFTSMC \cite{b7} & Fair & Exists & Medium/ & Medium/\\
		                 &  &  & High & Fast\\ \hline
	\end{tabular}
	\label{tab4}
\end{table}
Finally, we do a quantitative comparative analysis of the proposed control schemes. The performance indicators IAE, ITAE, $t_s$, and $e_{\Delta\mathrm{max}}$ are used for the quantitative comparison of the three controllers. Table \ref{tab3} shows a detailed quantitative performance comparison between the PI-SMC, the state feedback-based discrete-time SMC (DSMC), and the MROF-based discrete-time SMC (MROF-DSMC). Equation \eqref{eq83} is used to compute IAE, ITAE, and $e_{\Delta\mathrm{max}}$ for the three controllers. The system is perturbed by an external mismatched sinusoidal disturbance $d_i(z,t) = \sin(2\pi t)$ for the PI-SMC, and $d(k) = \sin(2\pi k\tau)$ for the two discrete-time SMCs, respectively. The settling time $t_s$, IAE, and ITAE are lowest for the PI-SMC and highest for the state feedback-based discrete-time SMC. Also, these three performance indicators for the MROF-based discrete-time SMC are significantly lower than the state feedback-based discrete-time SMC. The performance indicator $e_{\Delta\mathrm{max}}$ is highest for the state feedback-based discrete-time SMC and lowest for the MROF-based discrete-time SMC. Along with the settling time $t_s$ or the convergence rate, the indicator $e_{\Delta\mathrm{max}}$ impacts the total control effort and energy of the proposed controllers. Thus, the MROF-based discrete-time SMC requires a lower control effort and energy than the state feedback-based discrete-time SMC and almost a similar control effort and energy as the PI-SMC. The amplitude of chattering $\Delta e$ during the steady-state is maximum for the PI-SMC and lowest for the MROF-based discrete-time SMC. The phenomenon of chattering is discussed in more detail in the next section.

Furthermore, Table \ref{tab4} summarizes the qualitative comparison of the three proposed controllers along with the improved DPRL-I-SMC \cite{b8}, FOSMC \cite{b9}, and AFTSMC \cite{b7} control schemes. The high-frequency control input chattering present in the continuous-time SMCs is considerably reduced in the discrete-time counterparts, DSMC and MROF-DSMC, and the improved DPRL-I-SMC control scheme. The PI-SMC has the fastest convergence rate with zero steady-state error. A small steady-state error exists in the DSMC, FOSMC, and AFTSMC control schemes. The proposed MROF-DSMC control strategy possesses superior robustness characteristics and can effectively control the magnetic levitation system in the presence of mismatched uncertainties.

The simulation results demonstrate that the PI-SMC exhibits the best dynamic and steady-state performance with a fast convergence rate and zero steady-state error. Also, the MROF-based discrete-time SMC shows significantly better results than the state feedback-based discrete-time SMC in the presence of mismatched uncertainties. The MROF-based discrete-time SMC causes the system response to be close to that obtained by the PI-SMC or an equivalent continuous-time full-state feedback SMC. Thus, we can conclude that the MROF-based discrete-time SMC retains the invariance and robustness features of a continuous-time SMC to a greater extent than the state feedback-based discrete-time SMC.

\subsection{Chattering}\label{sec6subsec1}

The high-frequency oscillation of the control input about its reference value is called chattering. Since switching the control at an infinite rate is impossible, high-frequency chattering always occurs in a practical, continuous-time variable structure control system \cite{b13}. Chattering is significantly reduced in the discrete-time counterparts because the QSM motion of the state trajectory serves as a boundary layer for the controller. Figs. \ref{fig8}(a) and (b) show magnified versions of the control inputs for the state feedback-based discrete-time SMC and the MROF-based discrete-time SMC, respectively. For the state feedback-based discrete-time SMC, the voltage fluctuates between 8.2--8.36 V at a frequency of 0.625 Hz (large amplitude and low-frequency oscillations), whereas for the MROF-based discrete-time SMC, the voltage varies between 8.2711--8.2733 V at a frequency of 25 Hz (small amplitude and high-frequency oscillations). Thus, we can conclude that the frequency of chattering is reduced when the sampling rate is low as in the case of the state feedback-based discrete-time SMC but at the price of losing invariance. Whereas, with an increase in the sampling rate in the MROF-based discrete-time SMC, the frequency of chattering increases but at the same time robustness is enhanced. There is a trade-off between the chattering and robustness of the controller. Therefore, the MROF-based discrete-time SMC is designed such that it is not only robust to mismatched uncertainties but also the amplitude and frequency of chattering are low.

\section{Conclusion}\label{sec7}
In this paper, three types of sliding mode controllers have been proposed for a magnetic levitation system. A continuous-time PI-SMC has been developed using a new switching surface and a proportional plus power rate reaching law. Simulation results show that the PI-SMC is insensitive to deviations from the nominal operating point, is more robust than a feedback linearization controller, and outperforms the AFTSMC \cite{b7}, improved DPRL-I-SMC \cite{b8}, and FOSMC \cite{b9} control schemes in terms of the settling time. To reduce the chattering in the PI-SMC, a state feedback-based discrete-time SMC algorithm has been developed. However, because of the finite sampling frequency and QSM motion of the state trajectory, the disturbance rejection ability is compromised to some extent. 

To improve the robustness of the state feedback-based discrete-time SMC, we introduced a robust MROF-based discrete-time SMC strategy. With the MROF-based strategy, it is possible to realize the effect of a full-state feedback controller without incurring the complexity of a dynamic controller. Simulation results demonstrate that the MROF-based discrete-time SMC strategy can stabilize the magnetic levitation system with excellent dynamic and steady-state performance, fast convergence of the state trajectory to the QSM band, and reduced chatter in the presence of mismatched uncertainties. A quantitative and qualitative comparative analysis of the proposed control schemes is done using standard performance indicators. Using the Lyapunov stability theory, we have proved the asymptotic convergence of the system states to their desired equilibrium values for the proposed control schemes. Time delay is inevitably encountered for information transmission in the real-world application of practical magnetic levitation systems. The MROF-based strategy presented in this paper can be extended and applied to real-world time-delay and multirate systems \cite{b29,b30,b31}.

\end{document}